\documentclass[DIV=15, bibliography=totoc]{scrartcl}

\usepackage[a4paper,margin=2.25cm,bottom=2.5cm]{geometry}
\usepackage{float}
\usepackage[export]{adjustbox}
\usepackage[normalem]{ulem}
\usepackage{diagbox}
\usepackage{stmaryrd}
\usepackage{bm}
\usepackage{xstring}
\usepackage{flafter}
\usepackage{multirow}

\usepackage{xcolor} 
\definecolor{lightblue}{rgb}{0,0.5,1.0}
\definecolor{linkblue}{rgb}{0,0.1,0.6}
\definecolor{citegreen}{rgb}{0,0.4,0.0}
\definecolor{linkred}{rgb}{0.8,0,0.005}
\definecolor{mailviolet}{rgb}{0.3,0,0.35}
\definecolor{tumblue}{rgb}{0,0.396,0.741}
\definecolor{darkgreen}{rgb}{0,0.4,0} 
\definecolor{darkbrown}{rgb}{0.5, 0.396, 0.09}

\usepackage[hyperindex,colorlinks=true,linkcolor=linkblue,citecolor=citegreen,urlcolor=mailviolet,filecolor=linkred]{hyperref}
\usepackage[hyphenbreaks]{breakurl}
\usepackage{caption, subcaption}
\usepackage{booktabs}   
\usepackage{amsmath}
\usepackage{amsfonts}
\usepackage{lmodern}
\usepackage{graphicx}
\usepackage{listings}
\usepackage{todonotes}
\usepackage{placeins}
\usepackage{fancyhdr}
\usepackage{wrapfig}
\usepackage{algorithm}
\usepackage{algpseudocode}

\usepackage{enumitem}

\usepackage{siunitx}
\usepackage[square,comma,nonamebreak,compress,numbers]{natbib}

\usepackage{soulpos}
\usepackage{ulem}
\usepackage{authblk} 

\usepackage{pgfplots}
\usepackage{pgfplotstable}
\pgfplotsset{compat = newest}
\pgfplotsset{
  /pgfplots/my xbar legend/.style={
    /pgfplots/legend image code/.code={%
      \draw[##1,/tikz/.cd,bar width=.4em]
      plot coordinates { (1.2em, 1.0em)};}
}}
\usepackage{filecontents}
\usepackage{tikz}
\usepackage{tikzscale}
\usepackage{tikz-3dplot}

\usetikzlibrary{spy}
\usetikzlibrary{intersections}
\usetikzlibrary{arrows,shapes}
\usetikzlibrary{spy}
\usetikzlibrary{backgrounds}  
\usetikzlibrary{decorations}
\usetikzlibrary{decorations.markings}
\usetikzlibrary{positioning}
\usetikzlibrary{patterns}
\usetikzlibrary{calc} 
\usetikzlibrary{quotes} 
\usetikzlibrary{external} 
\usetikzlibrary{matrix}

\pgfplotscreateplotcyclelist{customCycleList}
{%
  {blue, mark=o},
  {red, mark=square},
  {darkgreen, mark=diamond},
  {cyan, mark=diamond},
  {darkbrown, mark=triangle*},
  {black, mark=pentagon},
}

\pgfplotsset{every axis/.append style= {
    cycle list name=customCycleList,
}}

\title{On the efficiency of explicit and semi-explicit immersed boundary finite element methods for wave propagation problems}

\newcommand*\samethanks[1][\value{footnote}]{\footnotemark[#1]}

\author[1]{Tim B{\"u}rchner\thanks{These two authors contributed equally to this work and share the first authorship.t}\thanks{\href{tim.buerchner@tum.de}{\texttt{tim.buerchner@tum.de}},
		Corresponding author}}
\author[2]{Lars Radtke\samethanks[1]\thanks{\href{lars.radtke@tuhh.de}{\texttt{lars.radtke@tuhh.de}},
		Corresponding author \\
Preprint submitted to Advances in Computational Science and Engineering}}
\author[3]{Philipp Kopp}
\author[3]{Stefan Kollmannsberger}
\author[1,4]{Ernst Rank}
\author[2]{Alexander Düster}

\affil[1]{Chair of Computational Modeling and Simulation, 
	Technische Universit\"at M\"unchen 
}
\affil[2]{Institute for Ship Structural Design and Analysis, 
	Technische Universit\"at Hamburg 
}
\affil[3]{Chair of Data Science in Civil Engineering, 
	Bauhaus-Universit\"at Weimar 
}
\affil[4]{Institute for Advanced Study, 
	Technische Universit\"at M\"unchen 
}

\date{}
\fancypagestyle{plain}{%
	\fancyhf{}%
	\fancyfoot[L]{
		\vspace{-0.0cm} 
} }

\newcommand{\tensor}[1]{\IfSubStr{ABCDEFGHIJKLMNOPQRSTUVWXYZ}{#1}
        {\mathbf{#1}}
        {\bm{#1}}}

\usepackage{xcolor}
\usepackage{comment} 
\usepackage[pagewise]{lineno}
\usepackage{amsmath}  
\usepackage{etoolbox} 

\newcommand*\linenomathpatch[1]{%
  \cspreto{#1}{\linenomath}%
  \cspreto{#1*}{\linenomath}%
  \csappto{end#1}{\endlinenomath}%
  \csappto{end#1*}{\endlinenomath}%
}

\linenomathpatch{equation}
\linenomathpatch{gather}
\linenomathpatch{multline}
\linenomathpatch{align}
\linenomathpatch{alignat}
\linenomathpatch{flalign}

\DeclareRobustCommand{\hg}[1]{{\sethlcolor{green}\hl{#1}}}

\begin{document}

\normalem \maketitle
\normalfont\fontsize{11}{13}\selectfont

\definecolor{myblue}{rgb}{0.0, 0.0, 1.0}
\DeclareRobustCommand{\hb}[1]{{\color{myblue}#1}} 
\definecolor{mygreen}{rgb}{0.0, 0.5, 0.0}
\DeclareRobustCommand{\hg}[1]{{\color{mygreen}#1}} 


\vspace{-1.5cm} \hrule

\newcommand{\lars}[1]{\textcolor{blue}{#1}}
\newcommand{\tim}[1]{\textcolor{green}{#1}}

\section*{Abstract}

Immersed boundary methods have attracted substantial interest in the last decades due to their potential for computations involving complex geometries. 
Often these cannot be efficiently discretized using boundary-fitted finite elements.
Immersed boundary methods provide a simple and fully automatic discretization based on Cartesian grids and tailored quadrature schemes that account for the geometric model.
It can thus be described independently of the grid, e.g., by image data obtained from computed tomography scans.
The drawback of such a discretization lies in the potentially small overlap between certain elements in the grid and the geometry.
These badly cut elements with small physical support pose a particular challenge for nonlinear and/or dynamic simulations.

In this work, we focus on problems in structural dynamics and acoustics and concentrate on solving them with explicit time-marching schemes.
In this context, badly cut elements can lead to unfeasibly small critical time step sizes. 
We investigate the performance of implicit-explicit time marching schemes and two stabilization methods developed in previous works as potential remedies.
While these have been studied before with regard to their effectiveness in increasing the critical time step size, their numerical efficiency has only been considered in terms of accuracy per degree of freedom.
In this paper, we evaluate the computation time required for a given accuracy, which depends not only on the number of degrees of freedom but also on the selected spatial discretization, the sparsity patterns of the system matrices, and the employed time-marching scheme. 

\vspace{0.25cm}
\noindent \textit{Keywords:} 
wave equation, explicit dynamics, finite cell method,  spectral element method, isogeometric analysis,
\vspace{0.25cm}


\section{Introduction}
{
\label{sec:sec1}

Immersed boundary finite element methods offer a versatile tool for the numerical analysis of complex geometries.
Instead of meshing complex geometries, they are embedded into a larger domain of simple geometry that can be discretized using simple meshes, e.g., Cartesian grids.
The actual geometry is taken into account using an indicator function that multiplies the integrands in the weak formulation of the underlying problem. To recover the original problem, the indicator function is set to one within the physical domain and to zero within the fictitious domain.

Over the last decades, a number of immersed boundary methods have been proposed that follow the idea of a fictitious domain approach.
In the present work, we utilize the \textit{finite cell method} (FCM), originally introduced in~\cite{parvizian2007, Duester2008}.
However, there are similar approaches, including \textit{CutFEM}~\cite{Burman2015}, \textit{cgFEM}~\cite{Nadal2013}, the \textit{shifted boundary method}~\cite{Main2018_1}, the \textit{cutDG method}~\cite{shoeder2020}, and several others.
The FCM combines the fictitious domain approach with high-order finite elements. To this end, the well-known Gaussian quadrature with $p+1$ points in every spatial dimension is employed for uncut cells, where $p$ is the polynomial degree of the shape functions. 
In cut cells, space tree methods are traditionally used to refine the quadrature grid and to accurately capture the discontinuity, also known as the immersed boundary.
While space trees offer a robust and fully automatic way to obtain a quadrature rule, the number of quadrature points is typically very high. 
To address this issue, moment fitting methods have been developed, see, e.g,~\cite{sudhakar2013,mueller2013,Legrain2021,Garhuom2022_2}. 
These methods effectively reduce the number of quadrature points, which is particularly important for nonlinear problems, where the system matrices must be computed repeatedly.

While immersed boundary methods allow to discretize very complex geometries in this way, cut elements which have only little support in the physical domain may lead to stability issues and poor conditioning of the system matrices. This, in turn, results in bad performance of iterative solvers as demonstrated in~\cite{sartorti2025} (in this special issue) for static problems and in very small critical time step sizes for explicit time integration schemes~\cite{Leidinger2020, radtke2024}.
In the context of FCM, the stability issues are typically mitigated by employing a small but positive value $\alpha$ instead of zero for the indicator function within the fictitious domain, which is often called $\alpha$-stabilization or material stabilization. 
In addition to the combination with high-order shape functions, this represents a key difference from other immersed boundary methods.
In~\cite{Schillinger2012}, the ideas of the FCM were coupled with isogeometric analysis (IGA), i.e., high-order shape functions based on B-splines. 
We denote this approach, also known as immersogeometric analysis, as IGA-FCM here and include it into our studies.

In recent years, the FCM has been employed more often in combination with explicit time integration schemes to solve dynamic problems. 
In particular, wave propagation problems represent a suitable application, as the time step requirements imposed by the modeled physical process are so stringent that the limitations imposed by badly cut cells in the discretization become acceptable.
In this context,~\cite{Joulaian2014} and~\cite{Duczek2014} combine the FCM with spectral shape functions, specifically Lagrange polynomials based on Gauss-Lobatto-Legendre~(GLL) points, and a quadrature based on these points. 
In uncut cells, this yields the spectral element method~\cite{Komatitsch1999, komatitsch2002}. 
Its combination with the ideas of the FCM is denoted as the spectral cell method (SCM).
The SEM is known for its enhanced spectral accuracy ($p$ times more accurate than the classical FEM) and high efficiency (due to a diagonal mass matrix by construction). 
The approach was examined for the scalar wave equation from a mathematical perspective in~\cite{ainsworth2009} and for structural dynamics from an engineering perspective in~\cite{radtke2021}.

Following the results in~\cite{Joulaian2014} and~\cite{Duczek2014}, several studies have addressed the possibility of applying lumping schemes, such as row-summing or diagonal scaling, to arrive at a fully diagonal mass matrix even for cut cells, where the interpolation points of the shape functions and the quadrature points cannot be directly matched.
To this end, we would like to mention the studies from~\cite{radtke2024} and the references therein, which demonstrate that it is generally less critical to apply such lumping schemes to discretizations based on Lagrange polynomials than to discretizations based on B-splines.
However, the critical time step size is more robust with respect to the cut position for B-splines than for Lagrange polynomials.
Generally speaking, a high-order convergent method yielding a fully diagonal mass matrix and, at the same time, an acceptably large critical time step size is yet unknown.
The issue of obtaining a diagonal mass matrix is the subject of several publications in the field of IGA, where dual bases are among the most promising approaches, see~\cite{Nguyen2023, held2024, hiemstra2023}.
In~\cite{radtke2024}, an overview of several alternative approaches is provided.

In the context of Lagrange shape functions, the critical time step size has been the subject of several recent publications. 
One remedy is the eigenvalue stabilization (EVS) method as introduced in~\cite{Eisentraeger2024}, which provides an effective alternative to the classical stabilization with a non-zero value~$\alpha$. 
The EVS method was originally introduced in~\cite{Loehnert2014} and was applied before in the context of the FCM for static problems in~\cite{Garhuom2022}.
It introduces a stabilization parameter $\epsilon$, which is proportional to the amount of stabilization. 
We refer to EVS as $\epsilon$-stabilization, in analogy to classical $\alpha$-stabilization.
Another potential remedy is the use of mixed or semi-explicit time integration schemes, which allow to treat the degrees of freedom associated with cut cells (and correspondingly non-diagonal mass matrix rows) with an implicit scheme. 
When combined with an explicit scheme for uncut cells, this results in an implicit-explicit (IMEX) time integration scheme as investigated in~\cite{May2017, Fassbender2024, Nicoli2024, Carson2025}.
It should be noted that stabilization methods targeting issues related to the critical time step size have also been developed in the context of other boundary fitted and immersed boundary methods. 
This includes the ghost-penalty method (see~\cite{Burman2022}) and cell agglomeration (see~\cite{shoeder2020}) as well as selective mass scaling (see~\cite{krauss2024}) and reciprocal mass matrices (see~\cite{Tkachuk20221}). 

In this work, we compare $\alpha$- and $\epsilon$-stabilization for the SCM and the IGA-FCM, when applied in combination with a consistent or partially lumped mass matrix and explicit, implicit, and IMEX time integration schemes.
A benchmark problem for scalar wave propagation is employed to investigate the computation times for the different discretization strategies and for varying stabilization parameters.
While previous studies have examined these strategies in terms of accuracy per degree of freedom, a comprehensive analysis of the computation time has yet to be conducted. 
A fair comparison of the computation time is feasible even when utilizing different code bases, as most of the computational effort is spent on time integration.
While the computation of the system matrices is typically highly implementation-dependent, the matrix vector products involved in the subsequent time integration can readily be realized using state-of-the art linear algebra frameworks, with the same implementation applied to all strategies.

The remainder of this work is structured as follows. In the next section, we introduce the basic equations for scalar wave propagation and their discretization in space and time using the selected strategies.
In Section~\ref{sec:numerical_investigations} we introduce the benchmark problem and present the results regarding the computational efficiency of the different methods.
Finally, we give concluding remarks and an outlook on future research in Section~\ref{sec:conclusion}.
}

\section{Methodology}
{
\label{sec:sec2}

In this section we briefly introduce the scalar wave equation as the considered model problem and its discretization using FCM. 
Afterwards we discuss the temporal discretization schemes and the stabilization methods.

\subsection{Partial differential equations}
We consider the scalar wave equation for the velocity potential $\Psi$
\begin{align}
    \rho \, \ddot{\Psi} - \nabla \cdot \left( \rho \, c^2 \, \nabla \Psi \right)  = \rho \, f \quad \text{ in } \Omega \times [0, T] \text{,}
    \label{eq:scalar_wave_equation}
\end{align}
where $\rho$ is the density, $c$ is the wave velocity and $f$ is a distributed load.
The acoustic pressure can be recovered as $p=\dot{\Psi}$, the acoustic particle velocity is obtained as $\bm{v}=\nabla \Psi$.
We consider Neumann and Dirichlet boundary conditions
\begin{align}
   \nabla \Psi \cdot \tensor{n} = \bar{v}_n \quad \text{ on } \Gamma^\mathrm{N}, \quad \text{ and } \quad 
   \Psi = \bar{\Psi} \quad \text{ on } \Gamma^\mathrm{D} \text{,}
   \label{eq:boundary_conditions}
\end{align}
where the unit normal vector $\tensor{n}$ on the domain's boundaries points outwards. 
We also specify initial states $\Psi(\tensor{x}, 0) = \Psi_0(\tensor{x})$ and $\dot{\Psi}(\tensor{x}, 0) = \dot{\Psi}_0(\tensor{x})$. 
In view of the fictitious domain method to be applied as explained in the next section, we do not divide Eq.~\eqref{eq:scalar_wave_equation} by the density $\rho$. 
This allows for an interpretation of the method as \textit{density scaling} (as opposed to wave speed scaling, see~\cite{Buerchner2023} for details}).

\subsection{Spatial Discretization}
In order to discretize \eqref{eq:scalar_wave_equation}, its weak form is derived by multiplication with a test function $\delta \Psi$ and integration by parts. After incorporating the Neumann boundary condition this yields
\begin{align}
    \int_\Omega  \rho \, \ddot{\Psi} \, \delta \Psi + \rho \, c^2 \, \nabla \Psi \cdot \nabla \delta\Psi \, \mathrm{d} \Omega 
    =
    \int_\Omega \rho \, f \, \delta \Psi \, \mathrm{d}\Omega + \int_{\Gamma^\mathrm{N}} \rho \, c^2 \, \bar{v}_n \, \delta \Psi \, \mathrm{d}\Gamma^\mathrm{N} \text{.}
    \label{eq:weak_form}
\end{align}
In the fictitious domain approach, the computational domain $\Omega$ is extended by a fictitious domain $\Omega^\mathrm{f}$ to obtain an extended domain $\Omega^\mathrm{e} = \Omega \cup \Omega^\mathrm{f}$, which has a simple geometry and can be easily meshed by a Cartesian grid.
Introducing the indicator function 
\begin{align}
    \alpha^\text{FCM}(\bm{x}) = \left\{ 
    \begin{array}{ll}
        1 & \text{ for } \bm{x} \in \Omega \\
        \alpha  &  \text{ else }
    \end{array}
    \right. ,
\end{align}
the weak form \eqref{eq:weak_form} can be written as
\begin{align}
    \int_{\Omega^\mathrm{e}}  \alpha^\text{FCM} \,  \rho \, \ddot{\Psi} \, \delta \Psi + \alpha^\text{FCM} \, \rho \, c^2 \, \nabla \Psi \cdot \nabla \delta\Psi \, \mathrm{d}\Omega^\mathrm{e} 
    =
    \int_{\Omega^\mathrm{e}} \alpha^\text{FCM} \,\rho \, f \, \delta \Psi \, \mathrm{d}\Omega^\mathrm{e} + \int_{\Gamma^\mathrm{N}} \alpha^\text{FCM} \, \rho \, c^2 \, \bar{v}_n \, \delta \Psi \, \mathrm{d}\Gamma^\mathrm{N} \text{.}
    \label{eq:weak_form_extended}
\end{align}
While any integral over $\Omega$ can be transformed into an integral over $\Omega^\mathrm{e}$ in this way, we would like to note that in a continuum mechanics setting, immersd boundary methods are often introduced as a scaling of physical parameters. 

We apply a classical finite element discretization for $\Psi$ and $\delta \Psi$ using the same set of tensor product shape functions.
This yields a system of ordinary differential equations
\begin{align}
    \mathbf{M} \, \ddot{\mathbf{\Psi}} + \mathbf{K} \, \mathbf{\Psi} = \mathbf{F} \label{eq:semi}
\end{align}
where $\mathbf{M}$ is the mass matrix, $\mathbf{K}$ is the stiffness matrix and $\mathbf{F}$ is the load vector. The vectors $\mathbf{\Psi}$ and $\ddot{\mathbf{\Psi}}$ contain the degrees of freedom for the velocity potential and its second time derivative, respectively.

We consider two different sets of shape functions, namely Lagrange polynomials and B-splines. The former are particularly attractive in explicit dynamics due to the possibility of nodal lumping. We apply a combination of Lagrange polynomials based on Gauß-Lobatto-Legendre (GLL) points, which are also used to establish the quadrature for the mass matrix in uncut cells. This effectively yields the spectral element method (SEM)~\cite{Komatitsch1999, komatitsch2002}. It is worth noting that the application of the GLL quadrature is only applied to the mass matrix, which is therefore under-integrated. Nevertheless, the spectral accuracy of the SEM is $p$ times higher than that of the FEM, as derived in~\cite{ainsworth2009} and investigated numerically in~\cite{radtke2021}.
In~\cite{Joulaian2014} and~\cite{Duczek2014}, the SEM was applied for the first time in combination with the fictitious domain ideas of the FCM and the resulting method was termed \textit{spectral cell method} (SCM).
Cut cells are integrated using a space tree partitioning and the classical quadrature based on Gauß-Legendre (GL) points in all leaf cells for all system matrices.
The mass matrices of cut cells can be lumped using diagonal scaling. However, this does not lead to efficient discretization schemes, as shown in~\cite{radtke2024,Fassbender2024}.

Using B-splines as shape functions leads to the isogeometric analysis (IGA) as introduced in~\cite{Hughes2005}, and investigated in the dynamic setting in~\cite{Cottrell2006}.
As mentioned in \cite{Cottrell2006}, and more recently shown in in~\cite{Buerchner2023b, radtke2024}, the high convergence rate of IGA is lost, if classical lumping schemes are applied. 
In addition to the loss of higher order convergence rates, the accuracy reduces by a large constant for single-step methods. 
As suggested in \cite{Messmer2022}, multi-rate time integration may offer a solution, but has not yet been investigated in detail.

\subsection{Temporal Discretization}
There is a variety of methods for temporal discretization of the semi-discrete system \eqref{eq:semi}. 
Despite its simplicity, the Newmark method~\cite{Newmark1959} is still one of the most widely used methods for integrating the wave equation in time. 
The time domain is defined as $\mathcal{T}=[0, T]$ and is divided into $N_\text{t}$ equal time intervals $[t_k, t_{k+1}]$ with time steps $\Delta t = \frac{T}{N_\mathrm{t}}$ for $k = 0, ..., N_\text{t}$, where $t_k = k \Delta t$ denotes a position in time. Introducing the weights $\beta$ and $\gamma$ for the displacement and velocity estimation, the Newmark method for the time step $k$ can be written as
\begin{align}
    &\tensor{\Psi}_{k+1} = \tensor{\Psi}_k + \Delta t \dot{\tensor{\Psi}}_k + \Delta t^2 \left( \left(\frac{1}{2}-\beta \right) \ddot{\tensor{\Psi}}_k + \beta \ddot{\tensor{\Psi}}_{k+1} \right) \\
    &\dot{\tensor{\Psi}}_{k+1} = \dot{\tensor{\Psi}}_k + \Delta t \left( \left(1 - \gamma\right) \ddot{\tensor{\Psi}}_k + \gamma \ddot{\tensor{\Psi}}_{k+1} \right),
\end{align}
where the acceleration vectors $\ddot{\tensor{\Psi}}_k$ and $\ddot{\tensor{\Psi}}_{k + 1}$ result from evaluating \eqref{eq:semi} at $t_k$ and $t_{k+1}$. For $\gamma=\frac{1}{2}$, the Newmark method is at least second-order accurate. The method can be implemented straightforwardly following the predictor-corrector scheme described in Algorithm~\ref{alg:newmark}.

\begin{algorithm}
\caption{Newmark method}\label{alg:newmark}
\begin{algorithmic}[H]
\Require $\tensor{\Psi}_0$, $\dot{\tensor{\Psi}}_0$, $\tensor{M}$, $\tensor{K}$, $\Delta t$, $\beta$, $\gamma$ 
\Ensure $\{\tensor{\Psi}_k\}_{k=0}^{N_\text{t}}$, $\{\dot{\tensor{\Psi}}_k\}_{k=0}^{N_\text{t}}$, $\{\ddot{\tensor{\Psi}}_k\}_{k=0}^{N_\text{t}}$
\State $\tensor{S} = \tensor{M} + \beta \Delta t^2 \tensor{K}$
\State $\ddot{\tensor{\Psi}}_0 = \tensor{M}^{-1} \left( \tensor{F}_k - \tensor{K} \tensor{\Psi}_0 \right)$
\For{$k = 0, ..., N-1$}
    \State $\tensor{\Psi}_\text{pred} = \tensor{\Psi}_k + \Delta t \dot{\tensor{\Psi}}_k + \left( \frac{1}{2} - \beta \right) \Delta t^2 \ddot{\tensor{\Psi}}_k$
    \State $\dot{\tensor{\Psi}}_\text{pred} = \dot{\tensor{\Psi}}_k + \left( 1 - \gamma \right) \Delta t \ddot{\tensor{\Psi}}_k$
    \State $\ddot{\tensor{\Psi}}_{k+1} = \tensor{S}^{-1}\left(\tensor{F}_{k+1} - \tensor{K}\tensor{\Psi}_\text{pred}\right)$
    \State $\dot{\tensor{\Psi}}_{k+1} = \dot{\tensor{\Psi}}_\text{pred} + \gamma \Delta t \ddot{\tensor{\Psi}}_{k+1}$
    \State $\tensor{\Psi}_{k+1} = \tensor{\Psi}_\text{pred} + \beta \Delta t^2 \ddot{\tensor{\Psi}}_{k+1}$
\EndFor 
\end{algorithmic}
\end{algorithm}

Setting $\beta = 0$ and $\gamma = \frac{1}{2}$ yields the explicit second-order central differences method~(CDM). After substituting the velocity vector, we can express the CDM can with a single equation
\begin{equation}
    \tensor{\Psi}_{k+1} = 2\tensor{\Psi}_k - \tensor{\Psi}_{k-1} + \Delta t^2 \tensor{M}^{-1} \left( \tensor{F}_k - \tensor{K} \tensor{\Psi}_k\right) \text{.}
    \label{eq:cdm}
\end{equation}
For a diagonal mass matrix $\tensor{M}$, the solution of \eqref{eq:cdm} becomes trivial. The CDM extrapolates the solution vector in time without needing to solve any system of equations. The CDM is conditionally stable with the critical time step size
\begin{equation}
    \Delta t^\text{crit} = \frac{2}{\sqrt{\lambda_\text{max}(\tensor{K}, \tensor{M})}}
\end{equation}
where $\lambda_\text{max}$ denotes the largest eigenvalue of the generalized eigenproblem of the semi-discretized system. By setting $\beta = \frac{1}{4}$ and $\gamma = \frac{1}{2}$, we obtain the implicit trapezoidal Newmark method, which remains unconditionally stable. An overview of time integration schemes used in combination with the SCM can be found in~\cite{Fassbender2023}.

\subsection{Lumping Schemes} 
\label{sec:lumping}

Nodal or also called optimal lumping is a scheme tailored to the spectral basis functions of SEM~\cite{Hughes2012, Duczek2019}.
Using the GLL points as interpolation points of the shape functions as well as quadrature points for the integration of the mass matrix inherently results in a diagonal mass matrix.
In the SCM, nodal lumping is applied in all uncut elements.
Accordingly, if we speak of lumping in the context of SCM, we refer to lumping of cut elements.

Independent of spectral basis functions, row-sum lumping is a simple and commonly used lumping scheme to obtain a diagonal mass matrix. 
In particular in low-order FEM, this approach improves the stability for explicit time-stepping methods and, therefore, is widely used~\cite{Voet2023}. 
To compute the diagonal entries of the lumped mass matrix, we sum all entries of the corresponding row. 
From an element point of view, we can write
\begin{align}
    &\tilde{m}_{ii}^e = \sum_j m_{ij}^e  \\
    &\tilde{m}_{ij}^e = 0 \quad \text{if } i \neq j \text{.}
\end{align}
In connection with immersed methods, the system may end up with negative diagonal entries, if the basis functions lack a nonnegative partition of unity property. The emerging singular or indefinite lumped mass matrix leads to unfeasible dynamic behavior. Regardless of the breakdown in $p$-convergence~\cite{Cottrell2006, Buerchner2023b}, row-sum lumping has experienced a revival in trimmed IGA, as the critical time step of cut elements becomes independent of the cut position~\cite{Leidinger2019, Messmer2022}. 

On the contrary, HRZ lumping~\cite{HRZ76} guarantees positive diagonal entries in the lumped mass matrix. Also known as diagonal scaling, the diagonal values of an element mass matrix are scaled to preserve the total mass of the element.
\begin{align}
    &\tilde{m}_{ii}^e = \frac{m^e}{\sum_j m_{jj}^e} m_{ii}^e  \\
    &\tilde{m}_{ij}^e = 0 \quad \text{if } i \neq j \quad     
\end{align}
where $m^e = \int_{\Omega^e} \rho d\Omega$ is the total mass of the element. 
In the context of SCM, it is often applied to lump cut elements' mass matrices.
Nevertheless, HRZ lumping also leads to a deterioration of the convergence rate~\cite{Duczek2019b, Fassbender2024}. For a review of lumping schemes applied to the SCM, refer to~\cite{Kelemen2021}. 

\subsection{Implicit-Explicit (IMEX) Time Integration}
In~\cite{Fassbender2024} it is proposed to use an implicit-explicit (IMEX) time integration scheme for a consistent version of the SCM. 
In the following, we briefly introduce the notation. 
In consistent SCM, the degrees of freedom can be partitioned according to the support of their basis functions. 
We reorder all degrees of freedom so that we can split $\tensor{\Psi}$ into 
\begin{equation}
     \tensor{\Psi} = \begin{bmatrix}
        \tensor{\Psi}^\text{d} \\ \tensor{\Psi}^\text{c}
    \end{bmatrix} \text{.}
\end{equation}
The vector $\tensor{\Psi}^\text{d}$ corresponds to the degrees of freedom that have support only in uncut elements contributing to a diagonal mass sub-matrix $\tensor{M}^\text{dd}$, while $\tensor{\Psi}^\text{c}$ represents the degrees of freedom that have support in at least one cut element contributing to a non-diagonal mass sub-matrix $\tensor{M}^\text{cc}$. After reordering, we can write
\begin{equation}
    \tensor{M} = \begin{bmatrix}
        \tensor{M}^\text{dd} & \tensor{0} \\
        \tensor{0} & \tensor{M}^\text{cc}
    \end{bmatrix} \text{.}
\end{equation}
Accordingly the stiffness matrix and the force vector are divided:
\begin{align}
    \tensor{K} &= \begin{bmatrix}
        \tensor{K}^\text{d} \\ \tensor{K}^\text{c}
    \end{bmatrix} = \begin{bmatrix}
        \tensor{K}^\text{dd} & \tensor{K}^\text{dc} \\
        \tensor{K}^\text{cd} & \tensor{K}^\text{cc}
    \end{bmatrix}, \\
    \tensor{F} &= \begin{bmatrix}
        \tensor{F}^\text{d} \\ \tensor{F}^\text{c}
    \end{bmatrix} \text{.}
\end{align}

The Newmark IMEX utilizes the CDM to integrate the uncut degrees of freedom explicitly in time. By contrast, the trapezoidal Newmark method integrates the cut degrees of freedom implicitly in time. Algorithm~\ref{alg:imex} illustrates the composed scheme. The Newmark IMEX inherits second-order accuracy from the two methods it is composed of and, in addition, its stability is determined solely by its explicit part, i.e., by the uncut elements~\cite{Hughes78a, Hughes78b}. 

\begin{algorithm}
\caption{Newmark IMEX}\label{alg:imex}
\begin{algorithmic}[H]
\Require $\tensor{\Psi}_0$, $\dot{\tensor{\Psi}}_0$, $\tensor{M}$, $\tensor{K}$, $\Delta t$, $\beta=\frac{1}{4}$, $\gamma=\frac{1}{2}$ 
\Ensure $\{\tensor{\Psi}\}_{k=0}^{N_\text{t}}$, $\{\dot{\tensor{\Psi}}\}_{k=0}^{N_\text{t}}$, $\{\ddot{\tensor{\Psi}}\}_{k=0}^{N_\text{t}}$
\State $\tensor{S} = \tensor{M}^\text{cc} + \beta \Delta t^2 \tensor{K}^\text{cc}$
\State $\ddot{\tensor{\Psi}}_0 = (\tensor{M}^\text{cc})^{-1} \left( \tensor{F}_k^\text{c} - \tensor{K}^\text{c} \tensor{\Psi}_0 \right)$
\For{$k = 0, ..., N-1$}
    \State Explicit step
    \State $\tensor{\Psi}_{k+1}^\text{d} = 2\tensor{\Psi}_k^\text{d} - \tensor{\Psi}_{k-1}^\text{d} + \Delta t^2 (\tensor{M}^\text{dd})^{-1} \left( \tensor{F}_k^\text{d} - \tensor{K}^\text{d} \tensor{\Psi}_k\right)$
    \State Implicit step
    \State $\tensor{\Psi}_\text{pred}^\text{d} = \tensor{\Psi}_{k+1}^\text{d}$
    \State $\tensor{\Psi}_\text{pred}^\text{c} = \tensor{\Psi}_k^\text{c} + \Delta t \dot{\tensor{\Psi}}_k^\text{c} + \left( \frac{1}{2} - \beta \right) \Delta t^2 \ddot{\tensor{\Psi}}_k^\text{c}$
    \State $\dot{\tensor{\Psi}}_\text{pred}^\text{c} = \dot{\tensor{\Psi}}_k^\text{c} + \left( 1 - \gamma \right) \Delta t \ddot{\tensor{\Psi}}_k^\text{c}$
    \State $\ddot{\tensor{\Psi}}_{k+1}^\text{c} = \tensor{S}^{-1}\left(\tensor{F}_{k+1}^\text{c} - \tensor{K}^\text{c} \tensor{\Psi}_\text{pred}\right)$
    \State $\dot{\tensor{\Psi}}_{k+1}^\text{c} = \dot{\tensor{\Psi}}_\text{pred}^\text{c} + \gamma \Delta t \ddot{\tensor{\Psi}}_{k+1}^\text{c}$
    \State $\tensor{\Psi}_{k+1}^\text{c} = \tensor{\Psi}_\text{pred}^\text{c} + \beta \Delta t^2 \ddot{\tensor{\Psi}}_{k+1}^\text{c}$
\EndFor 
\end{algorithmic}
\end{algorithm}

\subsection{Eigenvalue Stabilization}
The eigenvalue stabilization technique previously found application in the context of immersed boundary methods for non-linear static analyses, where it was utilized for the stiffness matrix~\cite{Garhuom2022}.
Here, we adopt the method proposed in~\cite{Eisentraeger2024}, which focuses on explicit dynamics.
To this end, we compute a stabilized element mass matrix $\mathbf{M}^e$ based on the \textit{original} element mass matrix $\mathbf{M}^\mathrm{o}$ and stabilization matrix $\mathbf{M}^\mathrm{s}$ as 
\begin{align}
    \mathbf{M}^e = \mathbf{M}^\mathrm{o} + \epsilon \, \mathbf{M}^\mathrm{s}
    \label{eq:epsilon_stabilization}
\end{align}
Introducing the \textit{full} element matrix $\mathbf{M}^\mathrm{f}$ that corresponds to an element that lies completely inside the physical domain $\Omega$, the $\alpha$-stabilization technique can be formulated as
\begin{align}
    \mathbf{M}^e = \mathbf{M}^\mathrm{o} + \alpha \, \left( \mathbf{M}^\mathrm{f} - \mathbf{M}^\mathrm{o} \right).
    \label{eq:alpha_stabilization}
\end{align}
Owing to the similarity between the structures of \eqref{eq:epsilon_stabilization} and \eqref{eq:alpha_stabilization}, the eigenvalue stabilization technique is referred to as $\epsilon$-stabilization as well.

In order to compute $\mathbf{M}^\mathrm{s}$, the unstabilized matrix $\mathbf{M}^\mathrm{o}$ is decomposed into a matrix $\mathbf{\Phi}$ consisting of its eigenvectors and a diagonal matrix $\mathbf{\Lambda}$ consisting of its eigenvalues. 
The eigenvalues are further classified as \textit{small} if $\lambda_i < f_\lambda \, \lambda_\mathrm{max}$ and collected in $\mathbf{\Lambda}^\mathrm{s}$ or as \textit{large} if $\lambda_i \geq f_\lambda \, \lambda_\mathrm{max}$  and collected in $\mathbf{\Lambda}^\mathrm{\ell}$, where $\lambda_\mathrm{max}$ is the largest eigenvalue and $f_\lambda$ is a given threshold. The decomposition can then be written as
\begin{align}
     \mathbf{M}^\mathrm{o} = \mathbf{\Phi} \, \mathbf{\Lambda} \, \mathbf{\Phi}^\mathrm{T}
     =
     \left[
\begin{array}{cc}
     \mathbf{\Phi}^\mathrm{s} &
     \mathbf{\Phi}^\mathrm{\ell} 
\end{array}
     \right]
          \left[
\begin{array}{cc}
     \mathbf{\Lambda}^\mathrm{s} & \bm{0} \\
     \bm{0} & \mathbf{\Lambda}^\mathrm{\ell} 
\end{array}
     \right]
     \left[
\begin{array}{cc}
     \mathbf{\Phi}^\mathrm{s} &
     \mathbf{\Phi}^\mathrm{\ell} 
\end{array}
     \right]^\mathrm{T}.
\end{align}
An unscaled stabilization matrix is constructed as
\begin{align}
    \tilde{\mathbf{M}}^\mathrm{s} = \mathbf{\Phi}^\mathrm{s}  \, {\mathbf{\Phi}^\mathrm{s}}^\mathrm{T} .
\end{align}
Removing small eigenvalues in the element mass matrices reduces the large eigenvalues of the generalized global eigenvalue problem and thus increases the critical time step size.
In order to make the stabilization matrix independent from the underlying material and element geometry, the final stabilization matrix is computed as
\begin{align}
    \mathbf{M}^\mathrm{s} = \frac{\mathrm{max}(\mathbf{M}^\mathrm{f})}{\mathrm{max}(\tilde{\mathbf{M}}^\mathrm{s})}\tilde{\mathbf{M}}^\mathrm{s}.
\end{align}
Therein, $\mathrm{max}(\mathbf{A})$ denotes the maximum element of a matrix $\mathbf{A}$.

\subsection{Selected discretization methods}
In summary, we consider two families of shape functions (Lagrange polynomials and B-splines) in combination with two stabilization methods ($\alpha$ and $\epsilon$) and two time integration schemes (the CDM and the trapezoidal rule), which may be combined to arrive at an IMEX scheme.
Since the stabilization methods are primarily used to reduce the time step size of the explicit CDM, which is associated with badly cut elements, we use them only when spatial discretizations are combined with CDM.
For spatial discretizations (B-splines or Lagrange polynomials) combined with IMEX or implicit Newmark-based temporal discretization only a very mild stabilization of $\alpha=10^{-12}$ is used. 
As introduced in~\cite{Joulaian2014} we denote the Lagrange-based discretization as SCM. The B-spline-based discretizations are denoted as IGA-FCM.
Table~\ref{tab:methods} gives an overview of all considered methods.
\begin{table}[H]
    \centering
    \caption{Discretizations considered in the efficiency studies.}
    \label{tab:methods}
    \renewcommand{\arraystretch}{1.3}
    \begin{tabular}{cccc}
        \toprule
               & CDM & implicit Newmark & IMEX\\ \hline
Lagrange       & SCM CDM  & not considered           & SCM IMEX \\ 
 & varying $\alpha$ and $\epsilon$ & & $\alpha=10^{-12}, \epsilon = 0$\\
        \midrule
B-splines      & IGA-FCM CDM & IGA-FCM NM & not considered \\ 
 & varying $\alpha, \epsilon = 0$ & $\alpha=10^{-12}, \epsilon = 0$ & \\ \bottomrule
    \end{tabular}
\end{table}

We would like to point out that all methods were also investigated in combination with a fully lumped mass matrix. However, the arising discretization schemes were not competitive in terms of efficiency (computation time for a given accuracy) as expected from the investigations in~\cite{Buerchner2023b, Fassbender2024, radtke2024} and shown in Figures \ref{fig:convergence_iga_alpha_lum} and \ref{fig:convergence_scm_alpha_lum}.

\section{Numerical Investigations}
\label{sec:numerical_investigations}

In this section, we investigate the efficiency of the various discretization methods in terms of computation time based on a benchmark problem.
We emphasize again that for the code parts which are most decisive for the computational times, i.e. the operations with the global system matrices, identical implementations based on optimized library functions are used, so that a fair comparison of the different methods can be made.
To solve linear equation systems in the time stepping, we use the direct solver \texttt{pardiso\_64} from Intel's oneMKL with matrix factorization and triangular substitution. 

\subsection{Benchmark problem}
\begin{figure}
    \centering
    \includegraphics[width=\textwidth]{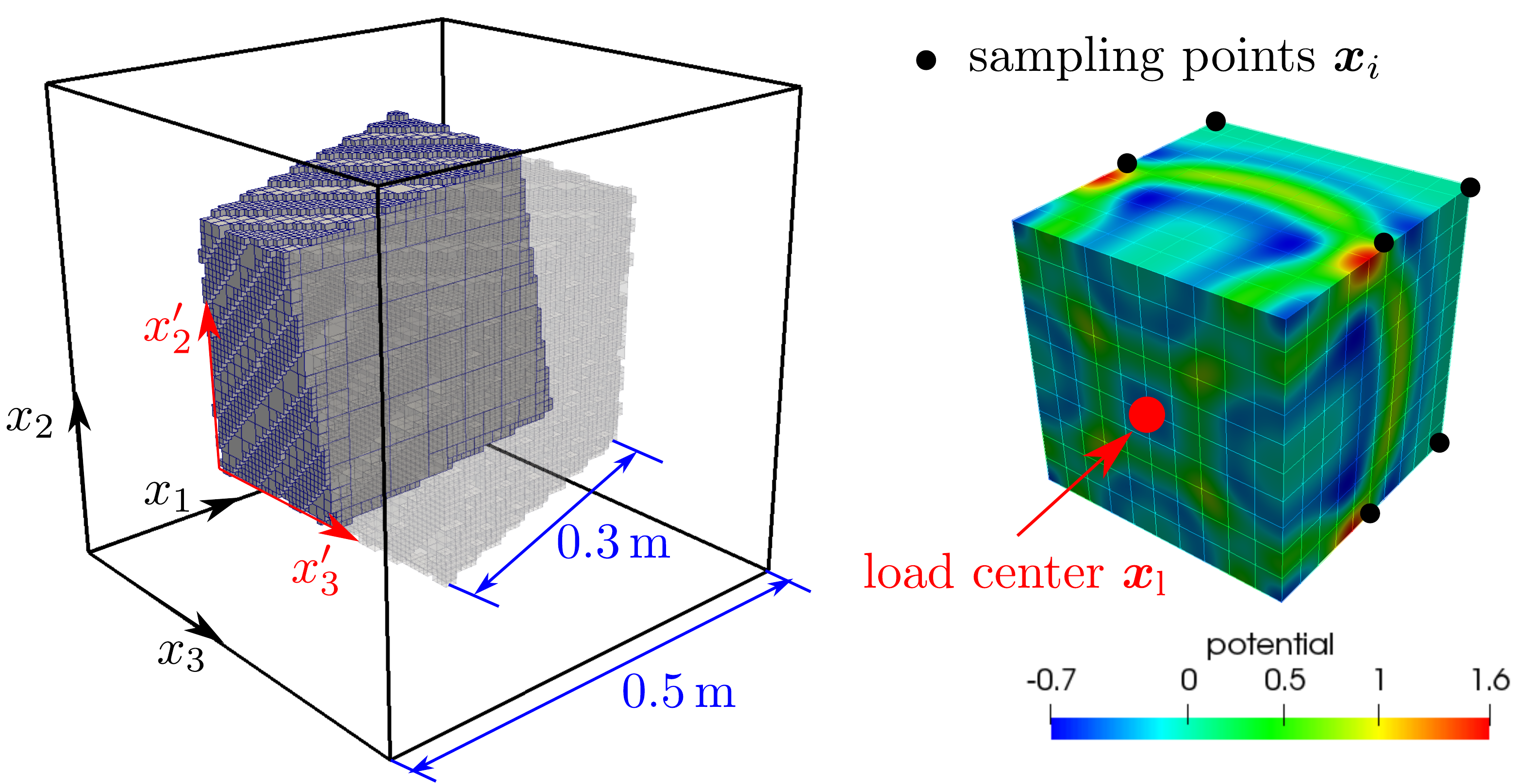}
    \caption{Rotated cube geometry with dimensions and indication of the octree subcells used for numerical integration~(left) and reference solution after $T=0.4\,$s~(right).}
    \label{fig:cube_overview}
\end{figure}

We choose a simple immersed cube geometry as the basis of our benchmark problem. 
A physical cube with edge length $l_\mathrm{p} = 0.3\,$m is embedded in a cubic extended domain with edge length $l_\mathrm{e} = 0.5\,$m.
While it can be easily meshed in a boundary-fitted manner, we rotate the physical cube inside the extended domain as shown in Figure~\ref{fig:cube_overview} (left).
The center of the two cubes coincides with the center of rotation. 
The transformation from the local to the global coordinate system is done using
\begin{align}
    \bm{x} = \bm{T} \, \bm{x}' \text{,}
\end{align}
where $\bm{T}$ is a rotation matrix corresponding to the chosen Cardan rotation angles.
For our study, we set the Cardan angles to $\phi = \SI{10}{\deg}$, $\theta = \SI{10}{\deg}$\, and $\psi = \SI{10}{\deg}$.

For a simulation time of $T=1$, we study the propagation of a wave initiated by a volumetric load induced at one of the face centers of the cube as indicated in  Figure~\ref{fig:cube_overview} (right).
In the coordinate system of the physical cube, the center of the excitation is $\bm{x}'_\mathrm{\ell}=[-\frac{l_\mathrm{p}}{2}, 0, 0]^\mathrm{T}$.
The volumetric source is distributed using a Gaussian function with standard deviation $\sigma_\mathrm{s} = 0.01$, i.e., 
\begin{equation}
    f_\mathrm{s}(\bm{x}')
    = 
    e^{- \frac{1}{2} \, d^2}, \quad \text{ with } d = \frac{\Vert \bm{x}' - \bm{x}'_\mathrm{\ell} \Vert_2}{\, \sigma_\mathrm{s}},
\end{equation}
where $\bm{x}'$ is the position vector in the local coordinate system.
The temporal excitation is described by a Ricker wavelet with the center frequency $f_\mathrm{e} = 10$ and a temporal shift $t_\mathrm{s} = \frac{2 \sqrt{6}}{\pi f_\mathrm{e}}$ (see~\cite{Ricker1943}):
\begin{equation}
    f_\mathrm{t}(t) 
    = 
    \left( 1 - 2 \, q^2 \right) \, e^{-q^2}, \quad \text{ with } q = \pi \, f_\mathrm{e} \, \left(t - t_\mathrm{s}\right).
    \label{eq:source_time}
\end{equation}

We observe the solution at 11 locations including the 4 corners where $x_1' = l$ and the 4 edge centers where $x_1' = \frac{l}{2}$ as indicated in Figure~\ref{fig:cube_overview} (right). Further, the source location $\bm{x}_l$ and the center of the cube width $x_1' = \frac{l}{2}$ and the face center with $x_1' = l$ are included as observer locations. Table~\ref{tab:observer_locations} gives all locations for the different types of observer locations. It is noted that due to symmetry, the exact solution only differs between locations with different types. This carries over to the solution of the boundary fitted discretizations as it preserves the symmetry. Due to the rotation, however, this does not hold for the immersed discretizations. Figure~\ref{fig:example_signal} illustrates this phenomenon for a selected discretization with $n^\mathrm{e}=10$ elements in each spatial direction and polynomial degree of $p=3$.
\begin{table}
    \centering
    \caption{Observer locations considered in the numerical accuracy studies.}
    \label{tab:observer_locations}
    \renewcommand{\arraystretch}{1.3}
    \begin{tabular}{ccccc}
        \toprule
index & $x_1$ & $x_2$ & $x_3$ & type \\ \hline
1     &  $0$ & $\frac{l}{2}$ & $\frac{l}{2}$ & source        \\ 
2     & $\frac{l}{2}$ & $\frac{l}{2}$ & $\frac{l}{2}$ & center \\ 
3     & $l$ & $\frac{l}{2}$ & $\frac{l}{2}$ & face \\ 
4,5,6,7     & $\frac{l}{2}$ & $\frac{l}{2} \pm \frac{l}{2}$ & $\frac{l}{2} \pm \frac{l}{2}$ & edge \\ 
8,9,19,11     & $l$ & $\frac{l}{2} \pm \frac{l}{2}$ & $\frac{l}{2} \pm \frac{l}{2}$ & corner \\ \bottomrule
    \end{tabular}
\end{table}
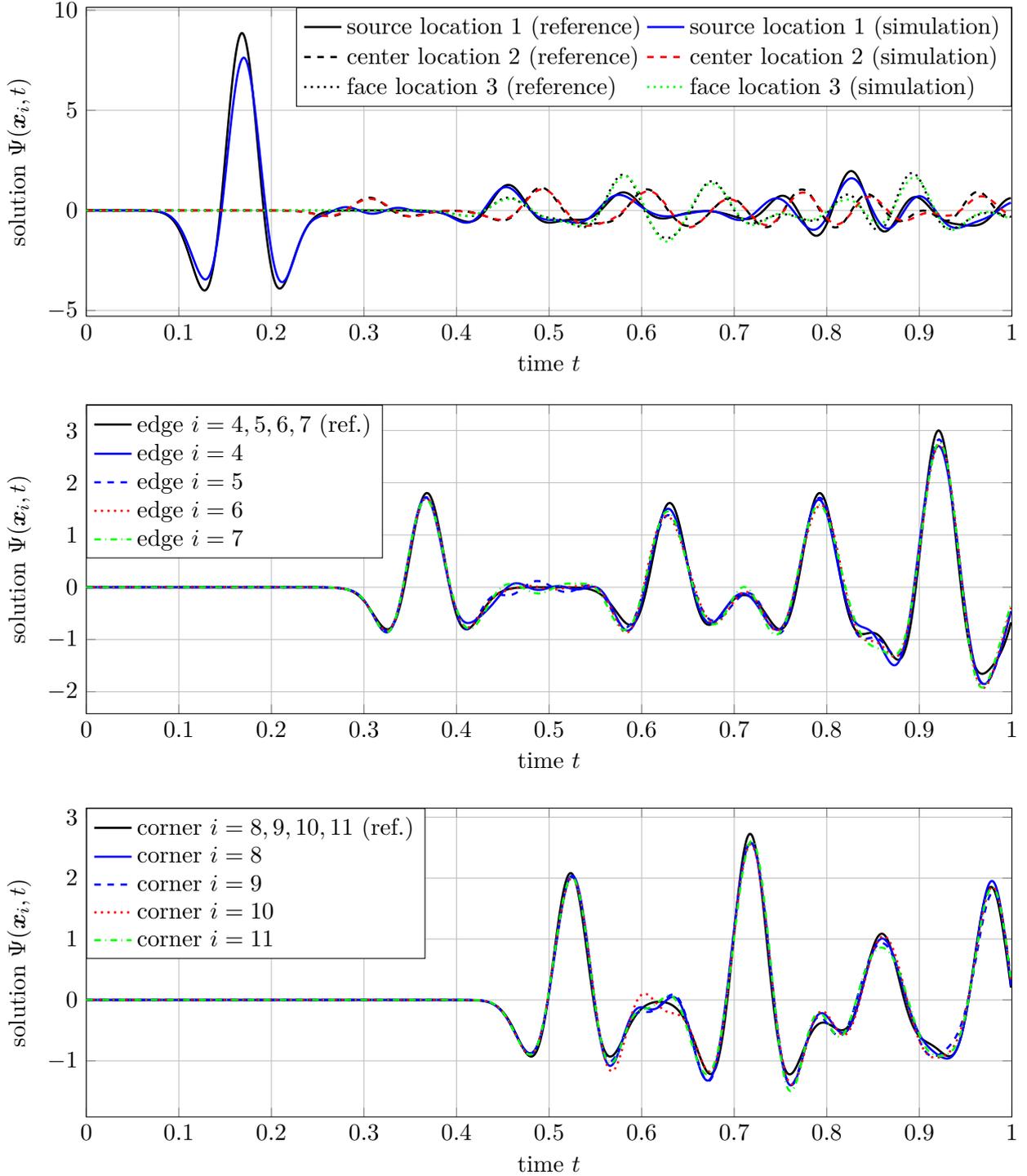
\begin{figure}[H]
	\centering
		\begin{tikzpicture}
	\begin{axis}[
		xmin = 0, xmax = 1,
		grid=major,
		width = \textwidth,
		height = 0.4\textwidth,
		xlabel = {time $t$},
		ylabel style={align=center}, 
        ylabel= {solution $\Psi(\bm{x}_i, t)$},
        legend columns = 2,
        legend cell align={left},
		legend style={at={(1,1)}, anchor=north east}]
		
		\addplot[black, mark = none, line width=1.0pt] file[] {pics/example_signal/resPoints_reference.dat};
		\addplot[blue, mark = none, line width=1.0pt] file[] {pics/example_signal/resPoints_n10_p3.dat};

  		\addplot[dashed, black, mark = none, line width=1.0pt] table[x index = 0, y index = 3] {pics/example_signal/resPoints_reference.dat};
		\addplot[dashed, red, mark = none, line width=1.0pt] table[x index = 0, y index = 3] {pics/example_signal/resPoints_n10_p3.dat};

        \addplot[dotted, black, mark = none, line width=1.0pt] table[x index = 0, y index = 5] {pics/example_signal/resPoints_reference.dat};
		\addplot[dotted, green, mark = none, line width=1.0pt] table[x index = 0, y index = 5] {pics/example_signal/resPoints_n10_p3.dat};

		\legend{source location 1 (reference), source location 1 (simulation), center location 2 (reference), center location 2 (simulation), face location 3 (reference), face location 3 (simulation)},
	\end{axis}
\end{tikzpicture}

\vspace{1em}

\begin{tikzpicture}
	\begin{axis}[
		xmin = 0, xmax = 1,
		grid=major,
		width = \textwidth,
		height = 0.4\textwidth,
		xlabel = {time $t$},
		ylabel style={align=center}, 
        ylabel= {solution $\Psi(\bm{x}_i, t)$},
        legend columns = 1,
        legend cell align={left},
		legend style={at={(0,1)}, anchor=north west}]
		
		\addplot[black, mark = none, line width=1.0pt] table[x index = 0, y index = 7] {pics/example_signal/resPoints_reference.dat};
  
		\addplot[blue, mark = none, line width=1.0pt] table[x index = 0, y index = 7] {pics/example_signal/resPoints_n10_p3.dat};

		\addplot[dashed, blue, mark = none, line width=1.0pt] table[x index = 0, y index = 11] {pics/example_signal/resPoints_n10_p3.dat};

		\addplot[dotted, red, mark = none, line width=1.0pt] table[x index = 0, y index = 15] {pics/example_signal/resPoints_n10_p3.dat};

		\addplot[dashdotted, green, mark = none, line width=1.0pt] table[x index = 0, y index = 19] {pics/example_signal/resPoints_n10_p3.dat};

		\legend{{edge $i=4,5,6,7$ (ref.)}, {edge $i=4$}, {edge $i=5$}, {edge $i=6$} , {edge $i=7$}}
  
	\end{axis}
\end{tikzpicture}

\vspace{1em}

\begin{tikzpicture}
	\begin{axis}[
		xmin = 0, xmax = 1,
		grid=major,
		width = \textwidth,
		height = 0.4\textwidth,
		xlabel = {time $t$},
		ylabel style={align=center}, 
        ylabel= {solution $\Psi(\bm{x}_i, t)$},
        legend columns = 1,
		legend cell align={left},
		legend style={at={(0,1)}, anchor=north west}]
		
		\addplot[black, mark = none, line width=1.0pt] table[x index = 0, y index = 9] {pics/example_signal/resPoints_reference.dat};
  
		\addplot[blue, mark = none, line width=1.0pt] table[x index = 0, y index = 9] {pics/example_signal/resPoints_n10_p3.dat};

		\addplot[dashed, blue, mark = none, line width=1.0pt] table[x index = 0, y index = 13] {pics/example_signal/resPoints_n10_p3.dat};

		\addplot[dotted, red, mark = none, line width=1.0pt] table[x index = 0, y index = 17] {pics/example_signal/resPoints_n10_p3.dat};

		\addplot[dashdotted, green, mark = none, line width=1.0pt] table[x index = 0, y index = 21] {pics/example_signal/resPoints_n10_p3.dat};

		\legend{{corner $i=8,9,10,11$ (ref.)}, {corner $i=8$}, {corner $i=9$}, {corner $i=10$} , {corner $i=11$}}
  
	\end{axis}
\end{tikzpicture}
    \caption{Signal at observer locations for a comparably coarse SCM discretization with $n^\mathrm{e}=10$, $p=3$, $\alpha=10^{-8}$ and $\epsilon = 0$.}
    \label{fig:example_signal}
\end{figure}

\subsection{Accuracy}
In order to select suitable discretizations for a computation time comparison, we first analyze the accuracy of a variety of candidate methods through convergence studies.
Based on these studies, we determine the most promising methods (in terms of accuracy per degree of freedom) that yield two desired levels of accuracy for the subsequent comparison.

\begin{figure}
	\centering
		\begin{tikzpicture}
	\begin{loglogaxis}[
		grid=major,
		width = 0.6\textwidth,
		height = 0.4\textwidth,
		xlabel = {$n^\mathrm{ts}$},
		ylabel style={align=center}, ylabel=$e^\mathrm{ts}$,
		legend style={at={(1,1)}, anchor=north east}]
		
		\addplot[black, mark = x, line width=1.0pt] file[] {pics/error_over_n_time_steps/cdm.dat};
		\addplot[blue, mark = *, line width=1.0pt] file[] {pics/error_over_n_time_steps/newmark.dat};
		
		\legend{CDM, Newmark},
	\end{loglogaxis}
\end{tikzpicture}
    \caption{Convergence of the solution for a boundary fitted SEM discretization with $n^\mathrm{e}=10$ and $p=5$ as the number of time steps is increased.}
    \label{fig:error_over_n_time_steps}
\end{figure}
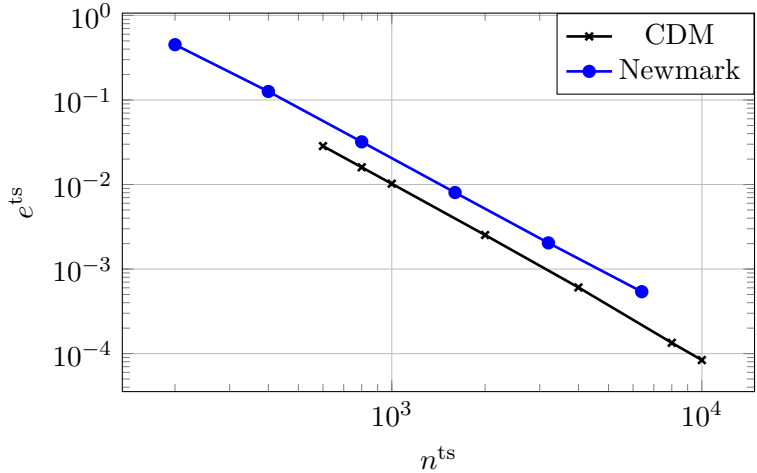

The accuracy is evaluated using a reference solution based on a very fine boundary fitted discretization and considering the solution $\Psi(\bm{x}_i, t)$ at the $n^\mathrm{p}=11$ observer locations $\bm{x}_i$ and $n^\mathrm{s}$ equidistant sampling points in time. Using
\begin{align}
    \mathbf{\Psi}^{i,n^\mathrm{s}} = \left[ \Psi(\bm{x}_i, t_1) ~ \Psi(\bm{x}_i, t_2) ~ \ldots ~ \Psi(\bm{x}_i, t_{n^\mathrm{s}}) \right]
\end{align}
the accuracy is determined based on the average relative error
\begin{align}
    e_{n^\mathrm{s}} = \frac{1}{n^\mathrm{p}} \, \sum_{i=1}^{n^\mathrm{p}} \, \frac{ \Vert  \mathbf{\Psi}^{i,n^\mathrm{s}} - \mathbf{\Psi}^{i,n^\mathrm{s}}_\mathrm{ref} \Vert_2  } { \Vert \mathbf{\Psi}^{i,n^\mathrm{s}}_\mathrm{ref} \Vert_2 }.
    \label{eq:error}
\end{align}
For the investigation of the accuracy of the time stepping schemes, we considered $e^\mathrm{ts} = e_{100}$. This makes it possible to compare simulations with different numbers of time steps $n^\mathrm{ts}$, as long as $n^\mathrm{ts}$ is a multiple of 100.
In the spatial convergence studies we consider $e^\mathrm{ss} = e_{10000}$, which includes all time steps of the respective simulations.
In order not to interfere with the error of the time integration scheme, $n^\mathrm{ts} = 10^5$ time steps were used in these spatial convergence studies resulting in a time step size of $\Delta t = 10^{-5}$. 
As shown in Figure~\ref{fig:error_over_n_time_steps}, this allows for an accuracy far beyond the desired accuracy levels in engineering applications. 
The reference solution for the temporal convergence study in Figure~\ref{fig:error_over_n_time_steps} was computed with a time step size of $\Delta t = 10^{-5}$ and $n^\mathrm{e}=10$ elements in each direction with a polynomial degree of $p=10$. The same reference solution is used in the following.

\FloatBarrier

\subsubsection{Spatial convergence}
We consider spatial convergence studies of the different discretization methods including a variation of the stabilization parameters $\alpha$ and $\epsilon$.
While a polynomial degree of $2 \leq p \leq 5$ was considered, we present the results in the form of $h$-convergence studies.

\begin{figure}[H]
	\centering
		\begin{tikzpicture}
	\begin{loglogaxis}[
		ymin = 3e-5, ymax = 1.1e-1,
		grid=major,
		width = 0.36\textwidth,
		height = 0.5\textwidth,
		xlabel = {$n^\mathrm{dof}$},
		ylabel style={align=center}, ylabel=$\Delta t^\mathrm{crit}$,
		legend style={at={(1,0)}, anchor=south west}]

  		\addplot[dashdotted, black, mark = square*, line width=1.0pt] 
table[x index = {0}, y index = {1}] 
{pics/convergence/scm_alpha_con/convergence_p2_alpha_1e-12_eps0.dat};
		\addplot[dashdotted, blue, mark = *, line width=1.0pt] 
table[x index = {0}, y index = {1}] 
{pics/convergence/scm_alpha_con/convergence_p3_alpha_1e-12_eps0.dat};
		\addplot[dashdotted, red, mark = square, line width=1.0pt] 
table[x index = {0}, y index = {1}] 
{pics/convergence/scm_alpha_con/convergence_p4_alpha_1e-12_eps0.dat};
        \addplot[dashdotted, green, mark = o, line width=1.0pt]
table[x index = {0}, y index = {1}] 
{pics/convergence/scm_alpha_con/convergence_p5_alpha_1e-12_eps0.dat};

		\addplot[black, mark = square*, line width=1.0pt] 
table[x index = {0}, y index = {1}] 
{pics/convergence/scm_alpha_con/convergence_p2_alpha_1e-8_eps0.dat};
		\addplot[blue, mark = *, line width=1.0pt] 
table[x index = {0}, y index = {1}] 
{pics/convergence/scm_alpha_con/convergence_p3_alpha_1e-8_eps0.dat};
		\addplot[red, mark = square, line width=1.0pt] 
table[x index = {0}, y index = {1}] 
{pics/convergence/scm_alpha_con/convergence_p4_alpha_1e-8_eps0.dat};
        \addplot[green, mark = o, line width=1.0pt]
table[x index = {0}, y index = {1}] 
{pics/convergence/scm_alpha_con/convergence_p5_alpha_1e-8_eps0_clean.dat};

          \addplot[dashed, mark options=solid, black, mark = square*, line width=1.0pt] 
table[x index = {0}, y index = {1}] 
{pics/convergence/scm_alpha_con/convergence_p2_alpha_1e-4_eps0.dat};
  		\addplot[dashed, mark options=solid, blue, mark = *, line width=1.0pt] 
table[x index = {0}, y index = {1}] 
{pics/convergence/scm_alpha_con/convergence_p3_alpha_1e-4_eps0.dat};
        \addplot[dashed, mark options=solid, red, mark = square, line width=1.0pt]
table[x index = {0}, y index = {1}] 
{pics/convergence/scm_alpha_con/convergence_p4_alpha_1e-4_eps0.dat};
        \addplot[dashed, mark options=solid, green, mark = o, line width=1.0pt] 
table[x index = {0}, y index = {1}] 
{pics/convergence/scm_alpha_con/convergence_p5_alpha_1e-4_eps0.dat};

        \addplot[dotted, mark options=solid, black, mark = square*, line width=1.0pt] 
table[x index = {0}, y index = {1}] 
{pics/convergence/scm_alpha_con/convergence_p2_alpha_1e-2_eps0.dat};
		 \addplot[dotted, mark options=solid, mark options=solid, blue, mark = *, line width=1.0pt] 
table[x index = {0}, y index = {1}] 
{pics/convergence/scm_alpha_con/convergence_p3_alpha_1e-2_eps0.dat};
        \addplot[dotted, mark options=solid, red, mark = square, line width=1.0pt] 
table[x index = {0}, y index = {1}] 
{pics/convergence/scm_alpha_con/convergence_p4_alpha_1e-2_eps0.dat};
        \addplot[dotted, mark options=solid, green, mark = o, line width=1.0pt] 
table[x index = {0}, y index = {1}] 
{pics/convergence/scm_alpha_con/convergence_p5_alpha_1e-2_eps0.dat};
  
		
	\end{loglogaxis}
\end{tikzpicture}
\begin{tikzpicture}
	\begin{loglogaxis}[
		ymin = 1.1e-4, ymax = 1.5,
		grid=major,
		width = 0.36\textwidth,
		height = 0.5\textwidth,
		xlabel = {$n^\mathrm{dof}$},
		ylabel style={align=center}, ylabel=$e^\mathrm{ss}$,
		legend style={at={(1.1,0)}, anchor=south west}]

      \addplot[dashdotted, black, mark = square*, line width=1.0pt] 
table[x index = {0}, y index = {2}] 
{pics/convergence/scm_alpha_con/convergence_p2_alpha_1e-12_eps0.dat};
        \addplot[dashdotted, blue, mark = *, line width=1.0pt] 
table[x index = {0}, y index = {2}] 
{pics/convergence/scm_alpha_con/convergence_p3_alpha_1e-12_eps0.dat};
        \addplot[dashdotted, red, mark = square, line width=1.0pt] 
table[x index = {0}, y index = {2}] 
{pics/convergence/scm_alpha_con/convergence_p4_alpha_1e-12_eps0.dat};
        \addplot[dashdotted, green, mark = o, line width=1.0pt] 
table[x index = {0}, y index = {2}] 
{pics/convergence/scm_alpha_con/convergence_p5_alpha_1e-12_eps0.dat};

		\addplot[black, mark = square*, line width=1.0pt] 
table[x index = {0}, y index = {2}] 
{pics/convergence/scm_alpha_con/convergence_p2_alpha_1e-8_eps0.dat};
        \addplot[blue, mark = *, line width=1.0pt] 
table[x index = {0}, y index = {2}] 
{pics/convergence/scm_alpha_con/convergence_p3_alpha_1e-8_eps0.dat};
        \addplot[red, mark = square, line width=1.0pt] 
table[x index = {0}, y index = {2}] 
{pics/convergence/scm_alpha_con/convergence_p4_alpha_1e-8_eps0.dat};
        \addplot[green, mark = o, line width=1.0pt] 
table[x index = {0}, y index = {2}] 
{pics/convergence/scm_alpha_con/convergence_p5_alpha_1e-8_eps0.dat};
  
		\addplot[dashed, mark options=solid, black, mark = square*, line width=1.0pt] 
table[x index = {0}, y index = {2}] 
{pics/convergence/scm_alpha_con/convergence_p2_alpha_1e-4_eps0.dat};
  		\addplot[dashed, mark options=solid, blue, mark = *, line width=1.0pt] 
table[x index = {0}, y index = {2}] 
{pics/convergence/scm_alpha_con/convergence_p3_alpha_1e-4_eps0.dat};
        \addplot[dashed, mark options=solid, mark options=solid, red, mark = square, line width=1.0pt] 
table[x index = {0}, y index = {2}] 
{pics/convergence/scm_alpha_con/convergence_p4_alpha_1e-4_eps0.dat};
        \addplot[dashed, mark options=solid, green, mark = o, line width=1.0pt] 
table[x index = {0}, y index = {2}] 
{pics/convergence/scm_alpha_con/convergence_p5_alpha_1e-4_eps0.dat};
        
		\addplot[dotted, mark options=solid, black, mark = square*, line width=1.0pt] 
table[x index = {0}, y index = {2}] 
{pics/convergence/scm_alpha_con/convergence_p2_alpha_1e-2_eps0.dat};
		\addplot[dotted, mark options=solid, blue, mark = *, line width=1.0pt] 
table[x index = {0}, y index = {2}] 
{pics/convergence/scm_alpha_con/convergence_p3_alpha_1e-2_eps0.dat};
        \addplot[dotted, mark options=solid, red, mark = square, line width=1.0pt] 
table[x index = {0}, y index = {2}] 
{pics/convergence/scm_alpha_con/convergence_p4_alpha_1e-2_eps0.dat};
        \addplot[dotted, mark options=solid, green, mark = o, line width=1.0pt] 
table[x index = {0}, y index = {2}] 
{pics/convergence/scm_alpha_con/convergence_p5_alpha_1e-2_eps0.dat};
  
		
		\legend{
  {$\alpha=10^{-12}, p=2$},{$\alpha=10^{-12}, p=3$}, {$\alpha=10^{-12}, p=4$}, {$\alpha=10^{-12}, p=5$}, 
  {$\alpha=10^{-8}, p=2$},{$\alpha=10^{-8}, p=3$}, {$\alpha=10^{-8}, p=4$}, {$\alpha=10^{-8}, p=5$}, 
  {$\alpha=10^{-4}, p=2$}, {$\alpha=10^{-4}, p=3$}, {$\alpha=10^{-4}, p=4$},  {$\alpha=10^{-4}, p=5$},
  {$\alpha=10^{-2}, p=2$}, {$\alpha=10^{-2}, p=3$}, {$\alpha=10^{-2}, p=4$}, {$\alpha=10^{-2}, p=5$}}
	\end{loglogaxis}
\end{tikzpicture}
    \caption{Critical time step size (left) and convergence behavior (right) of SCM with $\alpha$-stabilization.}
    \label{fig:convergence_scm_alpha_con}
\end{figure}
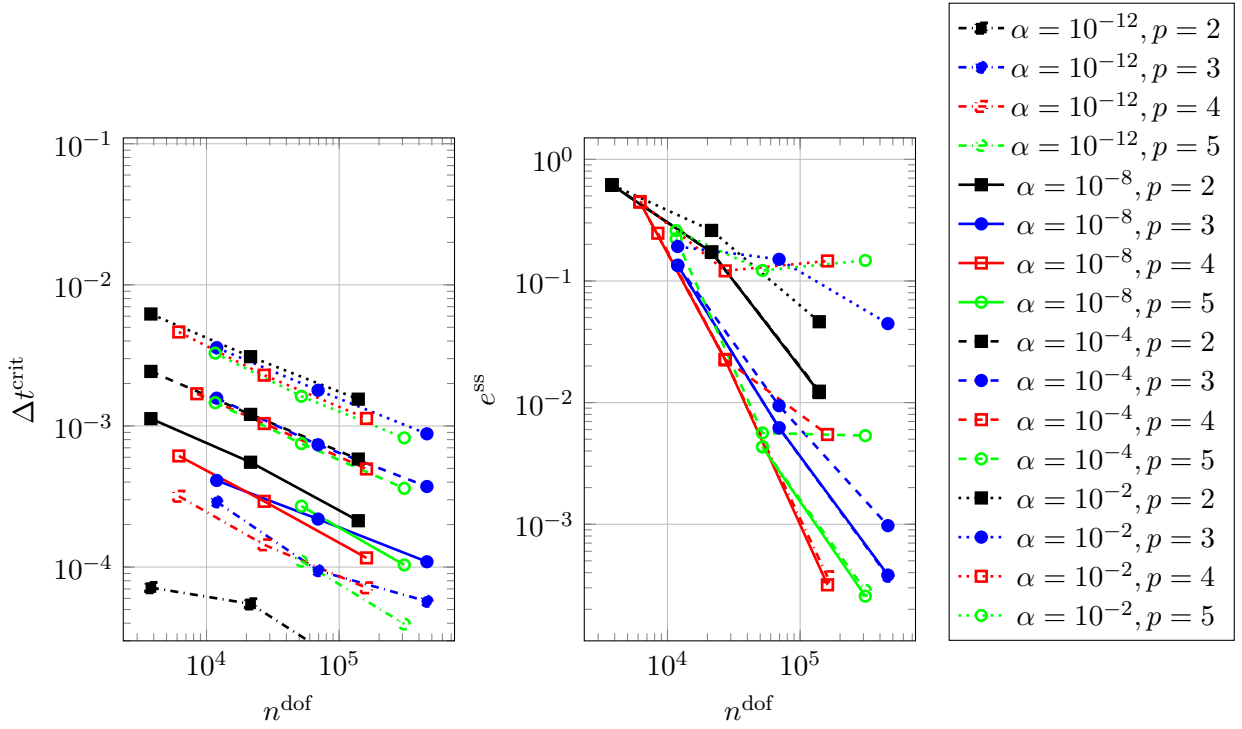

\paragraph{SCM with $\alpha$-stabilization}
Figure~\ref{fig:convergence_scm_alpha_con} shows the results obtained for the SCM with varying $\alpha$ and $\epsilon=0$.
The positive influence of large $\alpha$-values is clearly visible while at the same time, the accuracy is limited. In fact, the discretizations with the largest considered value of $\alpha=10^{-2}$ fail to deliver meaningful results with $e^\mathrm{ss}<0.05$, i.e., $5\%$. However, combining a moderate stabilization parameter $\alpha=10^{-4}$ with $p>2$ yields acceptable accuracies $e^\mathrm{ss}<0.01$, i.e., below $1\%$.
Accordingly, we only include SCM combined with CDM for $\alpha = 10^{-4}$ and $\alpha=10^{-8}$ into the later efficiency studies, and SCM combined with IMEX for $\alpha = 10^{-12}$.

\begin{figure}[H]
	\centering
		\begin{tikzpicture}
	\begin{loglogaxis}[
		ymin = 3e-5, ymax = 1.1e-1,
		grid=major,
		width = 0.36\textwidth,
		height = 0.5\textwidth,
		xlabel = {$n^\mathrm{dof}$},
		ylabel style={align=center}, ylabel=$\Delta t^\mathrm{crit}$,
		legend style={at={(1,0)}, anchor=south west}]
		
		\addplot[black, mark = square*, line width=1.0pt] 
table[x index = {0}, y index = {1}] 
{pics/convergence/scm_epsilon_con/convergence_p2_alpha_1e-12_eps_1e-6.dat};
        \addplot[blue, mark = *, line width=1.0pt] 
table[x index = {0}, y index = {1}] 
{pics/convergence/scm_epsilon_con/convergence_p3_alpha_1e-12_eps_1e-6.dat};
        \addplot[red, mark = square, line width=1.0pt] 
table[x index = {0}, y index = {1}] 
{pics/convergence/scm_epsilon_con/convergence_p4_alpha_1e-12_eps_1e-6.dat};
        \addplot[green, mark = o, line width=1.0pt] 
table[x index = {0}, y index = {1}] 
{pics/convergence/scm_epsilon_con/convergence_p5_alpha_1e-12_eps_1e-6.dat};
          
		\addplot[dashed, mark options=solid, black, mark = square*, line width=1.0pt] 
table[x index = {0}, y index = {1}] 
{pics/convergence/scm_epsilon_con/convergence_p2_alpha_1e-12_eps_1e-4.dat};
  		\addplot[dashed, mark options=solid, blue, mark = *, line width=1.0pt] 
table[x index = {0}, y index = {1}] 
{pics/convergence/scm_epsilon_con/convergence_p3_alpha_1e-12_eps_1e-4.dat};
        \addplot[dashed, mark options=solid, red, mark = square, line width=1.0pt] 
table[x index = {0}, y index = {1}] 
{pics/convergence/scm_epsilon_con/convergence_p4_alpha_1e-12_eps_1e-4.dat};
         \addplot[dashed, mark options=solid, green, mark = o, line width=1.0pt] 
table[x index = {0}, y index = {1}] 
{pics/convergence/scm_epsilon_con/convergence_p5_alpha_1e-12_eps_1e-4.dat};
         
		\addplot[dotted, mark options=solid, black, mark = square*, line width=1.0pt] 
table[x index = {0}, y index = {1}] 
{pics/convergence/scm_epsilon_con/convergence_p2_alpha_1e-12_eps_1e-2.dat};
		\addplot[dotted, mark options=solid, blue, mark = *, line width=1.0pt] 
table[x index = {0}, y index = {1}] 
{pics/convergence/scm_epsilon_con/convergence_p3_alpha_1e-12_eps_1e-2.dat};
        \addplot[dotted, mark options=solid, red, mark = square, line width=1.0pt] 
table[x index = {0}, y index = {1}] 
{pics/convergence/scm_epsilon_con/convergence_p4_alpha_1e-12_eps_1e-2.dat};
        \addplot[dotted, mark options=solid, green, mark = o, line width=1.0pt] 
table[x index = {0}, y index = {1}] 
{pics/convergence/scm_epsilon_con/convergence_p5_alpha_1e-12_eps_1e-2.dat};
		
	\end{loglogaxis}
\end{tikzpicture}
\begin{tikzpicture}
	\begin{loglogaxis}[
		ymin = 1.1e-4, ymax = 1.5,
		grid=major,
		width = 0.36\textwidth,
		height = 0.5\textwidth,
		xlabel = {$n^\mathrm{dof}$},
		ylabel style={align=center}, ylabel=$e^\mathrm{ss}$,
		legend style={at={(1.1,0)}, anchor=south west}]
		
		\addplot[black, mark = square*, line width=1.0pt] 
table[x index = {0}, y index = {2}] 
{pics/convergence/scm_epsilon_con/convergence_p2_alpha_1e-12_eps_1e-6.dat};
        \addplot[blue, mark = *, line width=1.0pt] 
table[x index = {0}, y index = {2}] 
{pics/convergence/scm_epsilon_con/convergence_p3_alpha_1e-12_eps_1e-6.dat};
        \addplot[red, mark = square, line width=1.0pt] 
table[x index = {0}, y index = {2}] 
{pics/convergence/scm_epsilon_con/convergence_p4_alpha_1e-12_eps_1e-6.dat};
        \addplot[green, mark = o, line width=1.0pt] 
table[x index = {0}, y index = {2}] 
{pics/convergence/scm_epsilon_con/convergence_p5_alpha_1e-12_eps_1e-6.dat};
          
		\addplot[dashed, mark options=solid, black, mark = square*, line width=1.0pt] 
table[x index = {0}, y index = {2}] 
{pics/convergence/scm_epsilon_con/convergence_p2_alpha_1e-12_eps_1e-4.dat};
  		\addplot[dashed, mark options=solid, blue, mark = *, line width=1.0pt] 
table[x index = {0}, y index = {2}] 
{pics/convergence/scm_epsilon_con/convergence_p3_alpha_1e-12_eps_1e-4.dat};
        \addplot[dashed, mark options=solid, red, mark = square, line width=1.0pt] 
table[x index = {0}, y index = {2}] 
{pics/convergence/scm_epsilon_con/convergence_p4_alpha_1e-12_eps_1e-4.dat};
         \addplot[dashed, mark options=solid, green, mark = o, line width=1.0pt] 
table[x index = {0}, y index = {2}] 
{pics/convergence/scm_epsilon_con/convergence_p5_alpha_1e-12_eps_1e-4.dat};
         
		\addplot[dotted, mark options=solid, black, mark = square*, line width=1.0pt] 
table[x index = {0}, y index = {2}] 
{pics/convergence/scm_epsilon_con/convergence_p2_alpha_1e-12_eps_1e-2.dat};
		\addplot[dotted, mark options=solid, blue, mark = *, line width=1.0pt] 
table[x index = {0}, y index = {2}] 
{pics/convergence/scm_epsilon_con/convergence_p3_alpha_1e-12_eps_1e-2.dat};
        \addplot[dotted, mark options=solid, red, mark = square, line width=1.0pt] 
table[x index = {0}, y index = {2}] 
{pics/convergence/scm_epsilon_con/convergence_p4_alpha_1e-12_eps_1e-2.dat};
        \addplot[dotted, mark options=solid, green, mark = o, line width=1.0pt] 
table[x index = {0}, y index = {2}] 
{pics/convergence/scm_epsilon_con/convergence_p5_alpha_1e-12_eps_1e-2.dat};
		
		\legend{
  {$\epsilon=10^{-6}, p=2$},{$\epsilon=10^{-6}, p=3$}, {$\epsilon=10^{-6}, p=4$}, {$\epsilon=10^{-6}, p=5$},
  {$\epsilon=10^{-4}, p=2$}, {$\epsilon=10^{-4}, p=3$}, {$\epsilon=10^{-4}, p=4$}, {$\epsilon=10^{-4}, p=5$},
  {$\epsilon=10^{-2}, p=2$}, {$\epsilon=10^{-2}, p=3$}, {$\epsilon=10^{-2}, p=4$}, {$\epsilon=10^{-2}, p=5$}}
	\end{loglogaxis}
\end{tikzpicture}
    \caption{Critical time step size (left) and convergence behavior (right) of SCM with $\epsilon$-stabilization.}
    \label{fig:convergence_scm_epsilon_con}
\end{figure}
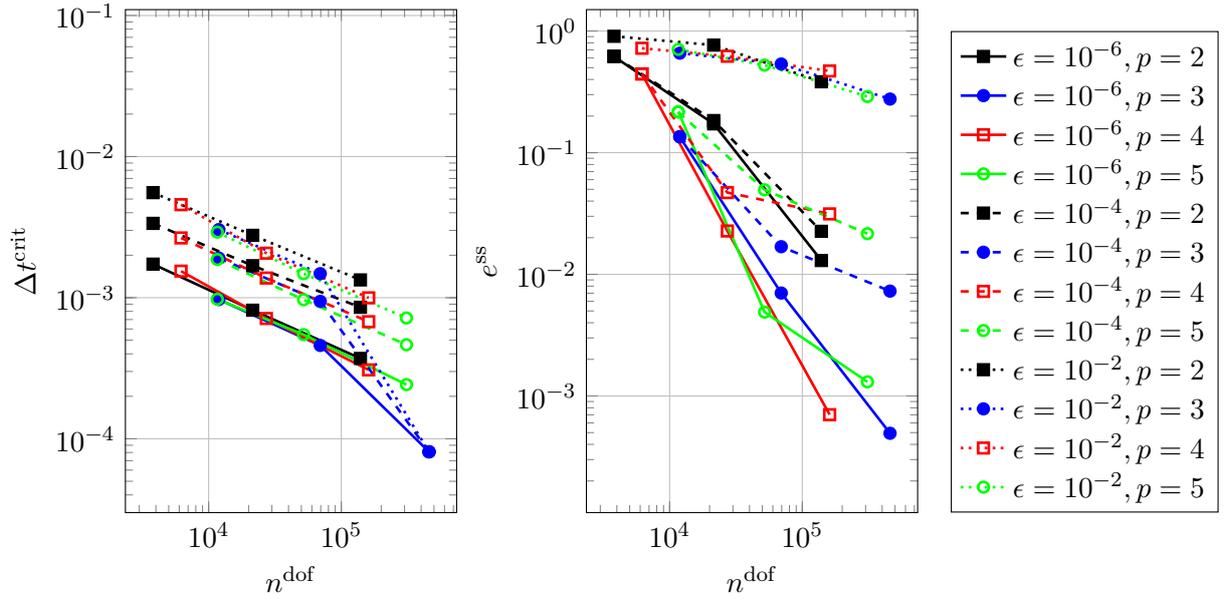

\paragraph{SCM with $\epsilon$-stabilization}
Figure~\ref{fig:convergence_scm_epsilon_con} shows the results for a study with varying $\epsilon$. Qualitatively, the effect is similar to $\alpha$-stabilization. Since the accuracy is severely degraded, we do not include the largest considered value of $\epsilon=10^{-2}$ in the later efficiency studies.
However, for the discretizations with $p=3$, the $\epsilon$-stabilization failed for some elements because the eigenvalue solver did not converge.
It is therefore recommended to always combine the $\epsilon$-stabilization with at least a small amount of $\alpha$-stabilization. Here we use $\alpha=10^{-12}$, which does not introduce a noticeable error for reasonably fine discretizations as confirmed by the accuracy studies in Section~\ref{sec:numerical_investigations}.
Further, we would like to point out the differences regarding the character of the error introduced by the $\epsilon$-stabilization compared to the $\alpha$-stabilization. For a similar reduction in the time step size, the error is larger for the $\epsilon$-stabilization. Accordingly, only $\epsilon=10^{-6}$ is considered for the later efficiency studies with a desired accuracy of $1\%$, i.e. $e^\mathrm{ss}=0.01$. For a desired accuracy of $5\%$, also $\epsilon=10^{-4}$ is considered. 

\begin{figure}[H]
	\centering
		\begin{tikzpicture}
\begin{loglogaxis}[
		ymin = 3e-5, ymax = 1.1e-1,
		grid=major,
		width = 0.36\textwidth,
		height = 0.5\textwidth,
		xlabel = {$n^\mathrm{dof}$},
		ylabel style={align=center}, ylabel=$\Delta t^\mathrm{crit}$,
		legend style={at={(1.0,0)}, anchor=south west}]
		
		\addplot[black, mark = square*, line width=1.0pt] table[x index = {0}, y index = {1}] {pics/convergence/iga_alpha_con/convergence_p2_alpha_1e-12_eps0.dat};
        \addplot[blue, mark = *, line width=1.0pt] table[x index = {0}, y index = {1}] {pics/convergence/iga_alpha_con/convergence_p3_alpha_1e-12_eps0.dat};
        \addplot[red, mark = square, line width=1.0pt] table[x index = {0}, y index = {1}] {pics/convergence/iga_alpha_con/convergence_p4_alpha_1e-12_eps0.dat};
        \addplot[green, mark = o, line width=1.0pt] table[x index = {0}, y index = {1}] {pics/convergence/iga_alpha_con/convergence_p5_alpha_1e-12_eps0.dat};
        
		\addplot[dashed, mark options=solid, black, mark = square*, line width=1.0pt] table[x index = {0}, y index = {1}] {pics/convergence/iga_alpha_con/convergence_p2_alpha_1e-8_eps0.dat};
		\addplot[dashed, mark options=solid, blue, mark = *, line width=1.0pt] table[x index = {0}, y index = {1}] {pics/convergence/iga_alpha_con/convergence_p3_alpha_1e-8_eps0.dat};
        \addplot[dashed, mark options=solid, red, mark = square, line width=1.0pt] table[x index = {0}, y index = {1}] {pics/convergence/iga_alpha_con/convergence_p4_alpha_1e-8_eps0.dat};
        \addplot[dashed, mark options=solid, green, mark = o, line width=1.0pt] table[x index = {0}, y index = {1}] {pics/convergence/iga_alpha_con/convergence_p5_alpha_1e-8_eps0.dat};

		\addplot[dotted, mark options=solid, black, mark = square*, line width=1.0pt] table[x index = {0}, y index = {1}] {pics/convergence/iga_alpha_con/convergence_p2_alpha_1e-4_eps0.dat};
		\addplot[dotted, mark options=solid, blue, mark = *, line width=1.0pt] table[x index = {0}, y index = {1}] {pics/convergence/iga_alpha_con/convergence_p3_alpha_1e-4_eps0.dat};
        \addplot[dotted, mark options=solid, red, mark = square, line width=1.0pt] table[x index = {0}, y index = {1}] {pics/convergence/iga_alpha_con/convergence_p4_alpha_1e-4_eps0.dat};
        \addplot[dotted, mark options=solid, green, mark = o, line width=1.0pt] table[x index = {0}, y index = {1}] {pics/convergence/iga_alpha_con/convergence_p5_alpha_1e-4_eps0.dat};

        \addplot[dashdotted, mark options=solid, black, mark = square*, line width=1.0pt] table[x index = {0}, y index = {1}] {pics/convergence/iga_alpha_con/convergence_p2_alpha_1e-2_eps0.dat};
		\addplot[dashdotted, mark options=solid, blue, mark = *, line width=1.0pt] table[x index = {0}, y index = {1}] {pics/convergence/iga_alpha_con/convergence_p3_alpha_1e-2_eps0.dat};
        \addplot[dashdotted, mark options=solid, red, mark = square, line width=1.0pt] table[x index = {0}, y index = {1}] {pics/convergence/iga_alpha_con/convergence_p4_alpha_1e-2_eps0.dat};
        \addplot[dashdotted, mark options=solid, green, mark = o, line width=1.0pt] table[x index = {0}, y index = {1}] {pics/convergence/iga_alpha_con/convergence_p5_alpha_1e-2_eps0.dat};
		
	\end{loglogaxis}
\end{tikzpicture}
\begin{tikzpicture}
	\begin{loglogaxis}[
		ymin = 1.1e-4, ymax = 1.5,
		grid=major,
		width = 0.36\textwidth,
		height = 0.5\textwidth,
		xlabel = {$n^\mathrm{dof}$},
		ylabel style={align=center}, ylabel=$e^\mathrm{ss}$,
		legend style={at={(1.1,0)}, anchor=south west}]
		
		\addplot[black, mark = square*, line width=1.0pt] table[x index = {0}, y index = {2}] {pics/convergence/iga_alpha_con/convergence_p2_alpha_1e-12_eps0.dat};
        \addplot[blue, mark = *, line width=1.0pt] table[x index = {0}, y index = {2}] {pics/convergence/iga_alpha_con/convergence_p3_alpha_1e-12_eps0.dat};
        \addplot[red, mark = square, line width=1.0pt] table[x index = {0}, y index = {2}] {pics/convergence/iga_alpha_con/convergence_p4_alpha_1e-12_eps0.dat};
        \addplot[green, mark = o, line width=1.0pt] table[x index = {0}, y index = {2}] {pics/convergence/iga_alpha_con/convergence_p5_alpha_1e-12_eps0.dat};
        
		\addplot[dashed, mark options=solid, black, mark = square*, line width=1.0pt] table[x index = {0}, y index = {2}] {pics/convergence/iga_alpha_con/convergence_p2_alpha_1e-8_eps0.dat};
		\addplot[dashed, mark options=solid, blue, mark = *, line width=1.0pt] table[x index = {0}, y index = {2}] {pics/convergence/iga_alpha_con/convergence_p3_alpha_1e-8_eps0.dat};
        \addplot[dashed, mark options=solid, red, mark = square, line width=1.0pt] table[x index = {0}, y index = {2}] {pics/convergence/iga_alpha_con/convergence_p4_alpha_1e-8_eps0.dat};
        \addplot[dashed, mark options=solid, green, mark = o, line width=1.0pt] table[x index = {0}, y index = {2}] {pics/convergence/iga_alpha_con/convergence_p5_alpha_1e-8_eps0.dat};

		\addplot[dotted, mark options=solid, black, mark = square*, line width=1.0pt] table[x index = {0}, y index = {2}] {pics/convergence/iga_alpha_con/convergence_p2_alpha_1e-4_eps0.dat};
		\addplot[dotted, mark options=solid, blue, mark = *, line width=1.0pt] table[x index = {0}, y index = {2}] {pics/convergence/iga_alpha_con/convergence_p3_alpha_1e-4_eps0.dat};
        \addplot[dotted, mark options=solid, red, mark = square, line width=1.0pt] table[x index = {0}, y index = {2}] {pics/convergence/iga_alpha_con/convergence_p4_alpha_1e-4_eps0.dat};
        \addplot[dotted, mark options=solid, green, mark = o, line width=1.0pt] table[x index = {0}, y index = {2}] {pics/convergence/iga_alpha_con/convergence_p5_alpha_1e-4_eps0.dat};

		\addplot[dashdotted, mark options=solid, black, mark = square*, line width=1.0pt] table[x index = {0}, y index = {2}] {pics/convergence/iga_alpha_con/convergence_p2_alpha_1e-2_eps0.dat};
		\addplot[dashdotted, mark options=solid, blue, mark = *, line width=1.0pt] table[x index = {0}, y index = {2}] {pics/convergence/iga_alpha_con/convergence_p3_alpha_1e-2_eps0.dat};
        \addplot[dashdotted, mark options=solid, red, mark = square, line width=1.0pt] table[x index = {0}, y index = {2}] {pics/convergence/iga_alpha_con/convergence_p4_alpha_1e-2_eps0.dat};
        \addplot[dashdotted, mark options=solid, green, mark = o, line width=1.0pt] table[x index = {0}, y index = {2}] {pics/convergence/iga_alpha_con/convergence_p5_alpha_1e-2_eps0.dat};

    \legend{
    {$\alpha=10^{-12}, p=2$},
    {$\alpha=10^{-12}, p=3$},
    {$\alpha=10^{-12}, p=4$}, 
    {$\alpha=10^{-12}, p=5$},
    {$\alpha=10^{-8}, p=2$},
    {$\alpha=10^{-8}, p=3$}, 
    {$\alpha=10^{-8}, p=4$}, 
    {$\alpha=10^{-8}, p=5$},
    {$\alpha=10^{-4}, p=2$},
    {$\alpha=10^{-4}, p=3$}, 
    {$\alpha=10^{-4}, p=4$}, 
    {$\alpha=10^{-4}, p=5$}, 
    {$\alpha=10^{-2}, p=2$},
    {$\alpha=10^{-2}, p=3$}, 
    {$\alpha=10^{-2}, p=4$}, 
    {$\alpha=10^{-2}, p=5$}
    }
	\end{loglogaxis}
\end{tikzpicture}
    \caption{Critical time step size (left) and convergence behavior (right) of IGA-FCM with $\alpha$-stabilization.}
    \label{fig:convergence_siga_alpha_con}
\end{figure}
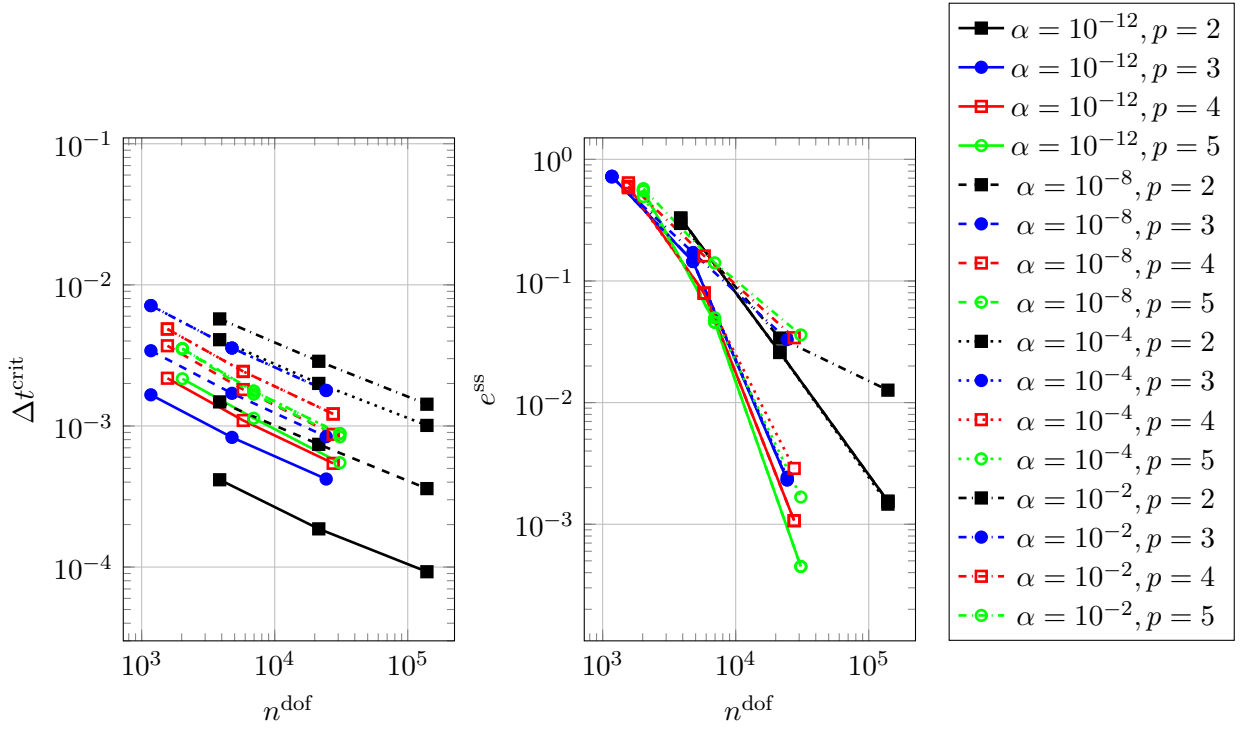

\paragraph{IGA-FCM}
For the IGA-FCM discretizations, we consider stabilized cases with $\epsilon=0$ and $\alpha$ varying between $\alpha=10^{-2}$ and $\alpha=10^{-12}$. 
Unlike the SCM discretizations the critical time step size is not as negatively affected by badly cut elements, respectively knot spans.
For a detailed study of this phenomenon, we refer to~\cite{radtke2024}.
The results of the convergence study are presented in Figure~\ref{fig:convergence_siga_alpha_con}. 
As expected, the accuracy per degree of freedom is much higher than for the SCM discretizations. 
However, the higher continuity also leads to more non-zeros in the sparse matrices. 
Further, the mass matrix is non-diagonal, even in uncut knot spans, while for the SCM, only the mass matrices of cut cells are non-diagonal.
Since the desired accuracy of 1\% is even reached with a stabilization of $\alpha=10^{-4}$, we include IGA-FCM for $\alpha=10^{-4}$ and $\alpha=10^{-8}$ in combination with CDM in the runtime study.
Additionally, we consider IGA-FCM with $\alpha=10^{-12}$ combined with the implicit Newmark method.

\paragraph{IGA-FCM and SCM with lumping}
For completeness, Fig.~\ref{fig:convergence_iga_alpha_lum} and \ref{fig:convergence_scm_alpha_lum} show the results for IGA-FCM and SCM, respectively, as obtained in combination with a lumped mass matrix.
Due to the arguments given in Sec.~\ref{sec:lumping}, we use classical row-summing for the IGA-FCM and diagonal scaling for the SCM.
The critical time step size of the lumped IGA-FCM is much larger than that of any other discretization (about one to two orders of magnitude), see Fig.~\ref{fig:convergence_iga_alpha_lum} (left). 
This is expected from the findings in~\cite{Leidinger2020, Messmer2022, radtke2024, bioli2024} and is achieved as well for $\alpha=0$.
However, the spatial error of the lumped IGA-FCM is much larger as well, see Fig.~\ref{fig:convergence_iga_alpha_lum} (right). 
In fact, no convergence is observed in the considered range of degrees of freedom.

For the lumped SCM, the picture is reversed. 
While it suffers from small critical time step sizes if not stabilized (see Fig.~\ref{fig:convergence_scm_alpha_lum} (left)), it shows at least slow convergence already in the considered preasymptotic range  (see Fig.~\ref{fig:convergence_scm_alpha_lum} (right)). However, the error is still unacceptably large so that none of the lumped discretizations are considered in the efficiency study.
\begin{figure}[H]
	\centering
		\begin{tikzpicture}
\begin{loglogaxis}[
		ymin = 3e-5, ymax = 1.1e-1,
		grid=major,
		width = 0.36\textwidth,
		height = 0.5\textwidth,
		xlabel = {$n^\mathrm{dof}$},
		ylabel style={align=center}, ylabel=$\Delta t^\mathrm{crit}$,
		legend style={at={(1.0,0)}, anchor=south west}]
		
		\addplot[black, mark = square*, line width=1.0pt] table[x index = {0}, y index = {1}] {pics/convergence/iga_alpha_lum/convergence_p2_alpha_1e-8_eps0.dat};
        \addplot[blue, mark = *, line width=1.0pt] table[x index = {0}, y index = {1}] {pics/convergence/iga_alpha_lum/convergence_p3_alpha_1e-8_eps0.dat};
        \addplot[red, mark = square, line width=1.0pt] table[x index = {0}, y index = {1}] {pics/convergence/iga_alpha_lum/convergence_p4_alpha_1e-8_eps0.dat};
        \addplot[green, mark = o, line width=1.0pt] table[x index = {0}, y index = {1}] {pics/convergence/iga_alpha_lum/convergence_p5_alpha_1e-8_eps0.dat};
        
		\addplot[dashed, mark options=solid, black, mark = square*, line width=1.0pt] table[x index = {0}, y index = {1}] {pics/convergence/iga_alpha_lum/convergence_p2_alpha_1e-12_eps0.dat};
		\addplot[dashed, mark options=solid, blue, mark = *, line width=1.0pt] table[x index = {0}, y index = {1}] {pics/convergence/iga_alpha_lum/convergence_p3_alpha_1e-12_eps0.dat};
        \addplot[dashed, mark options=solid, red, mark = square, line width=1.0pt] table[x index = {0}, y index = {1}] {pics/convergence/iga_alpha_lum/convergence_p4_alpha_1e-12_eps0.dat};
        \addplot[dashed, mark options=solid, green, mark = o, line width=1.0pt] table[x index = {0}, y index = {1}] {pics/convergence/iga_alpha_lum/convergence_p5_alpha_1e-12_eps0.dat};
		
	\end{loglogaxis}
\end{tikzpicture}
\begin{tikzpicture}
	\begin{loglogaxis}[
        ymin = 1.1e-4, ymax = 1.5,
		grid=major,
		width = 0.36\textwidth,
		height = 0.5\textwidth,
		xlabel = {$n^\mathrm{dof}$},
		ylabel style={align=center}, ylabel=$e^\mathrm{ss}$,
		legend style={at={(1.1,0)}, anchor=south west}]
		
		\addplot[black, mark = square*, line width=1.0pt] table[x index = {0}, y index = {2}] {pics/convergence/iga_alpha_lum/convergence_p2_alpha_1e-8_eps0.dat};
        \addplot[blue, mark = *, line width=1.0pt] table[x index = {0}, y index = {2}] {pics/convergence/iga_alpha_lum/convergence_p3_alpha_1e-8_eps0.dat};
        \addplot[red, mark = square, line width=1.0pt] table[x index = {0}, y index = {2}] {pics/convergence/iga_alpha_lum/convergence_p4_alpha_1e-8_eps0.dat};
        \addplot[green, mark = o, line width=1.0pt] table[x index = {0}, y index = {2}] {pics/convergence/iga_alpha_lum/convergence_p5_alpha_1e-8_eps0.dat};
        
		\addplot[dashed, mark options=solid, black, mark = square*, line width=1.0pt] table[x index = {0}, y index = {2}] {pics/convergence/iga_alpha_lum/convergence_p2_alpha_1e-12_eps0.dat};
		\addplot[dashed, mark options=solid, blue, mark = *, line width=1.0pt] table[x index = {0}, y index = {2}] {pics/convergence/iga_alpha_lum/convergence_p3_alpha_1e-12_eps0.dat};
        \addplot[dashed, mark options=solid, red, mark = square, line width=1.0pt] table[x index = {0}, y index = {2}] {pics/convergence/iga_alpha_lum/convergence_p4_alpha_1e-12_eps0.dat};
        \addplot[dashed, mark options=solid, green, mark = o, line width=1.0pt] table[x index = {0}, y index = {2}] {pics/convergence/iga_alpha_lum/convergence_p5_alpha_1e-12_eps0.dat};

		\legend{
  {$\alpha=10^{-8}, p=2$},{$\alpha=10^{-8}, p=3$}, {$\alpha=10^{-8}, p=4$}, {$\alpha=10^{-8}, p=5$}, 
  {$\alpha=10^{-12}, p=2$}, {$\alpha=10^{-12}, p=3$}, {$\alpha=10^{-12}, p=4$}, {$\alpha=10^{-12}, p=5$}}
	\end{loglogaxis}
\end{tikzpicture}
    \caption{Critical time step size (left) and convergence behavior (right) of IGA-FCM with $\alpha$-stabilization and a lumped mass matrix.}
    \label{fig:convergence_iga_alpha_lum}
\end{figure}
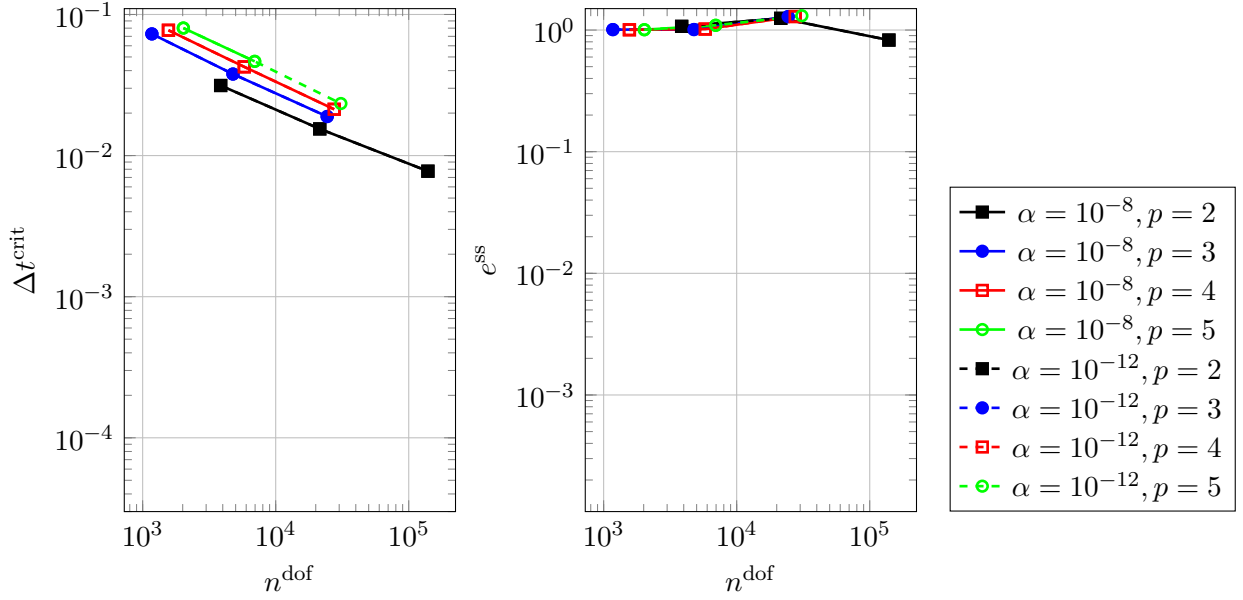
\begin{figure}[H]
	\centering
		\begin{tikzpicture}
	\begin{loglogaxis}[
		ymin = 3e-5, ymax = 1.1e-1,
		grid=major,
		width = 0.36\textwidth,
		height = 0.5\textwidth,
		xlabel = {$n^\mathrm{dof}$},
		ylabel style={align=center}, ylabel=$\Delta t^\mathrm{crit}$,
		legend style={at={(1,0)}, anchor=south west}]
		
		\addplot[black, mark = square*, line width=1.0pt] 
table[x index = {0}, y index = {1}] 
{pics/convergence/scm_alpha_lum/convergence_p2_alpha_1e-8_eps0.dat};
		\addplot[blue, mark = *, line width=1.0pt] 
table[x index = {0}, y index = {1}] 
{pics/convergence/scm_alpha_lum/convergence_p3_alpha_1e-8_eps0.dat};
		\addplot[red, mark = square, line width=1.0pt] 
table[x index = {0}, y index = {1}] 
{pics/convergence/scm_alpha_lum/convergence_p4_alpha_1e-8_eps0.dat};
        \addplot[green, mark = o, line width=1.0pt]
table[x index = {0}, y index = {1}] 
{pics/convergence/scm_alpha_lum/convergence_p5_alpha_1e-8_eps0.dat};

          \addplot[dashed, mark options=solid, black, mark = square*, line width=1.0pt] 
table[x index = {0}, y index = {1}] 
{pics/convergence/scm_alpha_lum/convergence_p2_alpha_1e-4_eps0.dat};
  		\addplot[dashed, mark options=solid, blue, mark = *, line width=1.0pt] 
table[x index = {0}, y index = {1}] 
{pics/convergence/scm_alpha_lum/convergence_p3_alpha_1e-4_eps0.dat};
        \addplot[dashed, mark options=solid, red, mark = square, line width=1.0pt]
table[x index = {0}, y index = {1}] 
{pics/convergence/scm_alpha_lum/convergence_p4_alpha_1e-4_eps0.dat};
        \addplot[dashed, mark options=solid, green, mark = o, line width=1.0pt] 
table[x index = {0}, y index = {1}] 
{pics/convergence/scm_alpha_lum/convergence_p5_alpha_1e-4_eps0.dat};

        \addplot[dotted, mark options=solid, black, mark = square*, line width=1.0pt] 
table[x index = {0}, y index = {1}] 
{pics/convergence/scm_alpha_lum/convergence_p2_alpha_1e-2_eps0.dat};
		 \addplot[dotted, mark options=solid, mark options=solid, blue, mark = *, line width=1.0pt] 
table[x index = {0}, y index = {1}] 
{pics/convergence/scm_alpha_lum/convergence_p3_alpha_1e-2_eps0.dat};
        \addplot[dotted, mark options=solid, red, mark = square, line width=1.0pt] 
table[x index = {0}, y index = {1}] 
{pics/convergence/scm_alpha_lum/convergence_p4_alpha_1e-2_eps0.dat};
        \addplot[dotted, mark options=solid, green, mark = o, line width=1.0pt] 
table[x index = {0}, y index = {1}] 
{pics/convergence/scm_alpha_lum/convergence_p5_alpha_1e-2_eps0.dat};
  
		
	\end{loglogaxis}
\end{tikzpicture}
\begin{tikzpicture}
	\begin{loglogaxis}[
		ymin = 1.1e-4, ymax = 1.5,
		grid=major,
		width = 0.36\textwidth,
		height = 0.5\textwidth,
		xlabel = {$n^\mathrm{dof}$},
		ylabel style={align=center}, ylabel=$e^\mathrm{ss}$,
		legend style={at={(1.1,0)}, anchor=south west}]
		
		\addplot[black, mark = square*, line width=1.0pt] 
table[x index = {0}, y index = {2}] 
{pics/convergence/scm_alpha_lum/convergence_p2_alpha_1e-8_eps0.dat};
        \addplot[blue, mark = *, line width=1.0pt] 
table[x index = {0}, y index = {2}] 
{pics/convergence/scm_alpha_lum/convergence_p3_alpha_1e-8_eps0.dat};
        \addplot[red, mark = square, line width=1.0pt] 
table[x index = {0}, y index = {2}] 
{pics/convergence/scm_alpha_lum/convergence_p4_alpha_1e-8_eps0.dat};
        \addplot[green, mark = o, line width=1.0pt] 
table[x index = {0}, y index = {2}] 
{pics/convergence/scm_alpha_lum/convergence_p5_alpha_1e-8_eps0.dat};
  
		\addplot[dashed, mark options=solid, black, mark = square*, line width=1.0pt] 
table[x index = {0}, y index = {2}] 
{pics/convergence/scm_alpha_lum/convergence_p2_alpha_1e-4_eps0.dat};
  		\addplot[dashed, mark options=solid, blue, mark = *, line width=1.0pt] 
table[x index = {0}, y index = {2}] 
{pics/convergence/scm_alpha_lum/convergence_p3_alpha_1e-4_eps0.dat};
        \addplot[dashed, mark options=solid, mark options=solid, red, mark = square, line width=1.0pt] 
table[x index = {0}, y index = {2}] 
{pics/convergence/scm_alpha_lum/convergence_p4_alpha_1e-4_eps0.dat};
        \addplot[dashed, mark options=solid, green, mark = o, line width=1.0pt] 
table[x index = {0}, y index = {2}] 
{pics/convergence/scm_alpha_lum/convergence_p5_alpha_1e-4_eps0.dat};
        
		\addplot[dotted, mark options=solid, black, mark = square*, line width=1.0pt] 
table[x index = {0}, y index = {2}] 
{pics/convergence/scm_alpha_lum/convergence_p2_alpha_1e-2_eps0.dat};
		\addplot[dotted, mark options=solid, blue, mark = *, line width=1.0pt] 
table[x index = {0}, y index = {2}] 
{pics/convergence/scm_alpha_lum/convergence_p3_alpha_1e-2_eps0.dat};
        \addplot[dotted, mark options=solid, red, mark = square, line width=1.0pt] 
table[x index = {0}, y index = {2}] 
{pics/convergence/scm_alpha_lum/convergence_p4_alpha_1e-2_eps0.dat};
        \addplot[dotted, mark options=solid, green, mark = o, line width=1.0pt] 
table[x index = {0}, y index = {2}] 
{pics/convergence/scm_alpha_lum/convergence_p5_alpha_1e-2_eps0.dat};
  
		
		\legend{
  {$\alpha=10^{-8}, p=2$},{$\alpha=10^{-8}, p=3$}, {$\alpha=10^{-8}, p=4$}, {$\alpha=10^{-8}, p=5$}, 
  {$\alpha=10^{-4}, p=2$}, {$\alpha=10^{-4}, p=3$}, {$\alpha=10^{-4}, p=4$},  {$\alpha=10^{-4}, p=5$},
  {$\alpha=10^{-2}, p=2$}, {$\alpha=10^{-2}, p=3$}, {$\alpha=10^{-2}, p=4$}, {$\alpha=10^{-2}, p=5$}}
	\end{loglogaxis}
\end{tikzpicture}
    \caption{Critical time step size (left) and convergence behavior (right) of SCM with $\alpha$-stabilization and a lumped mass matrix.}
    \label{fig:convergence_scm_alpha_lum}
\end{figure}
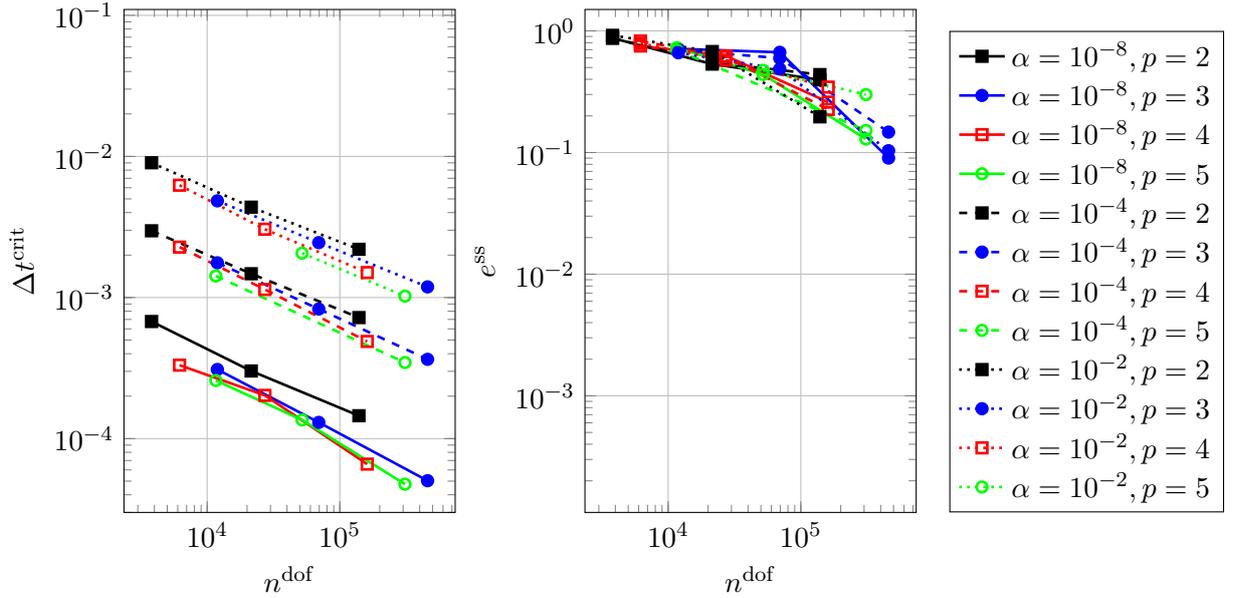

\subsubsection{Selection}
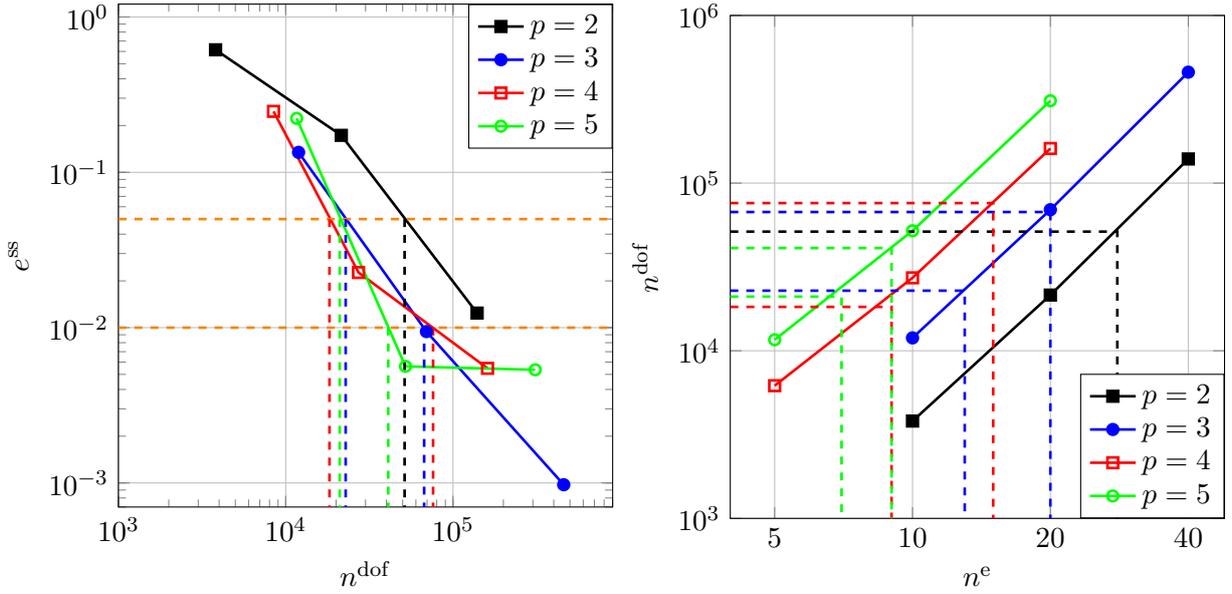
\begin{figure}[H]
	\centering
		\begin{tikzpicture}
	\begin{loglogaxis}[
		xmin = 1e3, 
        xmax = 9e5,
		ymin = 7e-4, 
		grid=major,
		width = 0.49\textwidth,
		height = 0.5\textwidth,
		xlabel = {$n^\mathrm{dof}$},
		ylabel style={align=center}, ylabel=$e^\mathrm{ss}$,
		legend style={at={(1,1)}, anchor=north east}]
		
		\addplot[black, mark = square*, line width=1.0pt] table[x index = {0}, y index = {2}] {pics/convergence/scm_alpha_con/convergence_p2_alpha_1e-4_eps0.dat};

        \addplot[blue, mark = *, line width=1.0pt] table[x index = {0}, y index = {2}] {pics/convergence/scm_alpha_con/convergence_p3_alpha_1e-4_eps0.dat};

        \addplot[red, mark = square, line width=1.0pt] table[x index = {0}, y index = {2}] {pics/convergence/scm_alpha_con/convergence_p4_alpha_1e-4_eps0.dat};

        \addplot[green, mark = o, line width=1.0pt] table[x index = {0}, y index = {2}] {pics/convergence/scm_alpha_con/convergence_p5_alpha_1e-4_eps0.dat};

        \addplot[dashed, orange, line width=1.0pt] coordinates {
            (1e3, 0.01)
            (9e5, 0.01)
        };

        \addplot[dashed, orange, line width=1.0pt] coordinates {
            (1e3, 0.05)
            (9e5, 0.05)
        };
                       
        \addplot[dashed, black, line width=1.0pt] coordinates {
            (51355, 0.05)
            (51355, 7e-4)
        };
        
        \addplot[dashed, blue, line width=1.0pt] coordinates {
            (67231, 0.01)
            (67231, 7e-4)
        };

        \addplot[dashed, blue, line width=1.0pt] coordinates {
            (22821, 0.05)
            (22821, 7e-4)
        };

        \addplot[dashed, red, line width=1.0pt] coordinates {
            (76029, 0.01) 
            (76029, 7e-4)
        };

        \addplot[dashed, red, line width=1.0pt] coordinates {
            (18268, 0.05) 
            (18268, 7e-4)
        };
               
        \addplot[dashed, green, line width=1.0pt] coordinates {
            (40989, 0.01) 
            (40989, 7e-4)
        };

        \addplot[dashed, green, line width=1.0pt] coordinates {
            (21025, 0.05) 
            (21025, 7e-4)
        };
        
		\legend{$p=2$, $p=3$, $p=4$, $p=5$},
        \title{SCM, $\varepsilon=10^{-4}$}
	\end{loglogaxis}
\end{tikzpicture}
\begin{tikzpicture}
	\begin{loglogaxis}[
		xmin = 4, xmax = 48,
		ymin = 1e3, ymax = 1e6,
		xtick={5, 10, 20, 40},
		xticklabels = {$5$, $10$, $20$, $40$},
		ytick={1000, 10000, 100000, 1000000},
		yticklabels={$10^3$, $10^4$, $10^5$, $10^6$},
	    yminorticks=false,
		grid=major,
		width = 0.49\textwidth,
		height = 0.5\textwidth,
		xlabel = {$n^\mathrm{e}$},
		ylabel style={align=center}, ylabel=$n^\mathrm{dof}$,
		legend style={at={(1,0)}, anchor=south east}]
		
		\addplot[black, mark = square*, line width=1.0pt] file[] {pics/dof_over_n/dof_over_n_p2.dat};
		\addplot[blue, mark = *, line width=1.0pt] file[] {pics/dof_over_n/dof_over_n_p3.dat};
		\addplot[red, mark = square, line width=1.0pt] file[] {pics/dof_over_n/dof_over_n_p4.dat};
		\addplot[green, mark = o, line width=1.0pt] file[] {pics/dof_over_n/dof_over_n_p5.dat};

        \addplot[dashed, black, line width=1.0pt] coordinates {
            (4, 51355)
            (28, 51355)
            (28, 1e3)
        };
        
        \addplot[dashed, blue, line width=1.0pt] coordinates {
            (4, 67231)
            (20, 67231)
            (20, 1e3)
        };

        \addplot[dashed, blue, line width=1.0pt] coordinates {
            (4, 22821)
            (13, 22821)
            (13, 1e3)
        };

        \addplot[dashed, red, line width=1.0pt] coordinates {
            (4, 76029) 
            (15, 76029) 
            (15, 1e3)
        };

        \addplot[dashed, red, line width=1.0pt] coordinates {
            (4, 18268)
            (9, 18268)
            (9, 1e3)
        };
               
        \addplot[dashed, green, line width=1.0pt] coordinates {
            (4, 40989) 
            (9, 40989) 
            (9, 1e3)
        };

        \addplot[dashed, green, line width=1.0pt] coordinates {
            (4, 21025)
            (7, 21025)
            (7, 1e3)
        };
        
		\legend{$p=2$, $p=3$, $p=4$, $p=5$},
	\end{loglogaxis}
\end{tikzpicture}
    \caption{Selection process shown for SCM discretization with $\alpha=10^{-4}$ and $\epsilon=0$. Step 1: find number of degrees of freedom $n^\mathrm{dof}$ from convergence graph (left). Step 2: select number of elements  $n^\mathrm{e}$ (right).}
    \label{fig:dof_over_n}
\end{figure}

From the convergence studies, we can choose suitable discretizations to be compared in terms of computation times. To begin with, we fix two levels of accuracy, $l_1 = 1\%$ and $l_2 = 5\%$.
We then use the linear interpolation in the double logarithmic plot and read off the number of degrees of freedom corresponding to the respective error levels (see Figure~\ref{fig:dof_over_n} (left)). 
Then, using the relationship between the number of degrees of freedom and the number of elements $n^\mathrm{e}$ as shown in Figure~\ref{fig:dof_over_n} (right), we select an appropriate $n^\mathrm{e}$ for the comparison that ensures an accuracy of the desired level. 
Note that the relationship in Figure~\ref{fig:dof_over_n} cannot be determined analytically in a trivial way because elements that are located completely outside the cube domain are disregarded from the discretization.

Table~\ref{tab:timing_discretizations} summarizes all discretizations selected for the comparison of computation times.
As mentioned before, some discretizations are not considered because they failed to deliver accurate results.
The desired accuracy is either not reached at all during the convergence study or it could only be reached with an unacceptable large number of degrees of freedom (larger than $10^{5}$ or even $10^{6}$).
These discretizations are excluded as indicated by the missing value for $n^\mathrm{e}$.
Again, we would like to point out that discretizations based on a lumped mass matrix for cut elements would require even larger (up to several orders of magnitude) number of degrees of freedom for the desired accuracy levels.
\begin{table}
    \centering
    \caption{Discretizations considered in the timing studies. The number of degrees of freedom (dof) refer to the number read of during the selection and in parenthesis the actual number of dof resulting from the chosen number of elements $n^\mathrm{e}$.}
    \label{tab:timing_discretizations}
    \begin{tabular}{lccccccc}
        \toprule
               & & \multicolumn{3} {c} {$e < 5\%$} & \multicolumn{3} {c} {$e < 1\%$} \\
        method & $p$ & dof & $n^\mathrm{e}$ & $\Delta t^\mathrm{crit} \cdot 10^{3}$ & dof & $n^\mathrm{e}$ & $\Delta t^\mathrm{crit} \cdot 10^3$ \\ \hline
\multirow{4}{*}{\begin{minipage}[]{100pt}
SCM con. \\ $\alpha=10^{-12}$, $\epsilon=0$ IMEX
\end{minipage}} 
         & $2$ & 51348 (52353) & 28 & 5.96237 & $>10^5$ & - & \\
         & $3$ & 21081 (22816) & 13 & 5.84696 & 51266 (52375) & 18 & 4.17564 \\
         & $4$ & 18033 (21109) & 9 & 5.17179 & 38615 (43065) & 12 & 3.75073 \\
         & $5$ & 20341 (22706) & 7 & 4.43017 & 39162 (40176) & 9 & 3.42847 \\ \hline
\multirow{4}{*}{\begin{minipage}[]{100pt}
SCM con. \\ $\alpha=10^{-8}$, $\epsilon=0$ \\ CDM
\end{minipage}} 
         & $2$ & 51178 (52353) & 28 & 0.43236 & $>10^5$ & - &  \\
         & $3$ & 20998 (22816) & 13 & 0.38268 & 52726 (52375) & 18 & 0.25533 \\
         & $4$ & 18272 (21109) &  9 & 0.43601 & 38096 (43065) & 12 & 0.24549 \\
         & $5$ & 18646 (22706) &  7 & 0.41463 & 36673 (40176) & 9 & 0.31423 \\ \hline
\multirow{4}{*}{\begin{minipage}[]{100pt}
SCM con. \\ $\alpha=10^{-4}$, $\epsilon=0$ \\ CDM
\end{minipage}} 
         & $2$ & 51355 (52353) & 28 & 0.86921 & $>10^5$ & - & - \\
         & $3$ & 22821 (22816) & 13 & 1.29185 & 67231 (69391) & 20 & 0.735765 \\
         & $4$ & 18268 (21109) & 9 & 1.16246 & 76029 (75085) & 15 & 0.698116 \\
         & $5$ & 21025 (22706) & 7 & 1.00886 & 40989 (40176) & 9 & 0.906885 \\ \hline
\multirow{4}{*}{\begin{minipage}[]{100pt}
SCM con. \\ $\alpha=10^{-12}$, $\epsilon=10^{-6}$\\ CDM
\end{minipage}} 
         & $2$ & 52726 (52353) & 28 & 0.45652 & $>10^5$ & - & - \\
         & $3$ & 21398 (22816) & 13 & 0.74380 & 55906 (60472) & 19 & 0.51546 \\
         & $4$ & 18162 (21109) & 9 & 0.87312 & 41472 (43065) & 12 & 0.63272 \\
         & $5$ & 20902 (22706) & 7 & 0.74016 & 38885 (40176) & 9 & 0.56692 \\ \hline
\multirow{4}{*}{\begin{minipage}[]{100pt}
SCM con. \\ $\alpha=10^{-12}$, $\epsilon=10^{-4}$  \\ CDM
\end{minipage}} 
         & $2$ & 68623 (69083) & 31 & 0.41570 & $>10^5$ & - & - \\
         & $3$ & 27606 (27793) & 14 & 1.34215 & 227948 (226108) & 31 & 0.19435 \\
         & $4$ & 26035 (27233) & 10 & 1.37805 & - & - & - \\
         & $5$ & 52112 (67086) & 11 & 0.88394 & - & - & - \\ \hline
\multirow{4}{*}{\begin{minipage}[]{100pt}
IGA-FCM con. \\ $\alpha=10^{-12}$, $\epsilon=0$ \\ Newmark
\end{minipage}} 
         & $2$ & 13634 (14130) & 34 & 0.23916 & 39923 (40751) & 51 & 0.14488 \\
         & $3$ &  7137  (7221) & 24 & 0.68262 & 13433 (14079) & 32 & 0.48951 \\
         & $4$ &  6770  (7116) & 22 & 0.96601 & 12160 (11984) & 28 & 0.71641 \\
         & $5$ &  6633  (6951) & 20 & 1.13415 & - & - & - \\ \hline
\multirow{4}{*}{\begin{minipage}[]{100pt}
IGA-FCM con. \\ $\alpha=10^{-8}$, $\epsilon=0$ \\ CDM
\end{minipage}} 
         & $2$ & 13811 (14130) & 34 & 0.79411 & 40457 (40751) & 51 & 0.54669 \\
         & $3$ &  7290  (7904) & 25 & 1.11988 & 13699 (14079) & 32 & 1.04079 \\
         & $4$ &  6890  (7116) & 22 & 1.59176 & 12288 (12967) & 29 & 1.22720 \\
         & $5$ &  6716  (6951) & 20 & 1.68044 & - & - & - \\ \hline
\multirow{4}{*}{\begin{minipage}[]{100pt}
IGA-FCM con. \\ $\alpha=10^{-4}$, $\epsilon=0$ \\ CDM
\end{minipage}} 
         & $2$ & 13884 (14130) & 34 & 2.36090 & 40715 (40751) & 51 & 1.57227 \\
         & $3$ &  7333  (7904) & 25 & 2.63871 & 13887 (14079) & 32 & 2.23040 \\
         & $4$ &  7227  (7829) & 23 & 2.11658 & 15340 (16250) & 32 & 1.52100 \\
         & $5$ &  7010  (7678) & 21 & 1.68287 & - & - & - \\ \bottomrule
    \end{tabular}
\end{table}

\clearpage
\newpage
\subsection{Computation times}
In this section, we present the comparison of the selected methods in terms of computation times. 
To this end, two different hardware setups are considered.
We denote as \textit{Hardware 1} an Intel Core i9-9900K processor at 3.60GHz and as \textit{Hardware 2} an Intel Xeon E5-2640 v4 at 2.40GHz.
The simulation procedure can be divided into the following stages.
\setlist{nolistsep}
\begin{itemize}[noitemsep]
    \item \textbf{System setup}. This stage includes the integration of the system matrices as well as their manipulation, i.e., lumping and stabilization.
    \item \textbf{Factorization}. In this stage, the system matrix is factorized, i.e., $\mathbf{M}$ for computations solely based on the CDM and a combination of $\mathbf{M}$ and $\mathbf{K}$ for computation involving the Newmark method.
    \item \textbf{Time stepping.} This stage includes the computations done in every time step. We separate them into two parts:
    \begin{itemize}
        \item \textbf{Right-hand side computation}. This part includes the evaluation of the right-hand side, which is dominated by the product $\mathbf{K}\, \mathbf{u}_{i}$.
        \item \textbf{Backward insertion}. This part includes the computation of the degrees of freedom $\mathbf{u}_{i+1}$ based on the factorized system matrix and the previously computed right-hand-side.
    \end{itemize}
\end{itemize}
The following studies compare the runtimes of the methods for ten independent single-threaded runs, taking into account only the factorization and the time stepping stages.
All simulations are performed using a time step size as large as possible to guarantee the desired accuracy level as well as stability. 
To this end, we set
\begin{align}
    \Delta t = \min(0.9 \, \Delta t^\mathrm{crit},
    \Delta t^\mathrm{max}).
\end{align}
For a desired accuracy of 1\% we set $\Delta t^\mathrm{max}=10^{-3}$ and for a desired accuracy of 5\% we set $\Delta t^\mathrm{max}=0.00\bar{2}$ which ensures the accuracy according to Figure~\ref{fig:error_over_n_time_steps}.
This corresponds to a number of time steps of $n^\mathrm{ts}=1000$ and $n^\mathrm{ts}=450$, respectively.
For variants including the implicit Newmark method, i.e., SCM IMEX and IGA-FCM Newmark, we choose $\Delta t^\mathrm{max} = 0.015385$ corresponding to $n^\mathrm{ts} = 650$ for 5\% accuracy, and $\Delta t^\mathrm{max} = 0.0068966$ corresponding to $n^\mathrm{ts} = 1450$ for 1\% accuracy. 

\subsubsection{Results}

Figure~\ref{fig:runtimes_tim} shows the measured runtimes for all chosen methods on Hardware 1, Figure~\ref{fig:runtimes_lars} on Hardware 2. 
The polynomial degrees are coded according to the colors.
The average of the ten runs is shown as bars.
In addition, all individual runtimes are plotted as dots to visualize how the runtimes cluster for each method. 
The left sides depict the studies with 5\% accuracy, the right sides with 1\% accuracy.
Configurations not included in the study were assigned a value of zero. 
It is noted that especially on Hardware 1 the individual runtimes for a method are very similar across the ten runs, resulting in an overlay of the individual dots.
\begin{figure}[htb!]
    \captionsetup[subfigure]{oneside,margin={2.5cm,0cm}}
	\begin{subfigure}[t]{0.4\textwidth}
		\centering
		\begin{tikzpicture}
	\begin{axis}[
    xlabel={Computation Time $[s]$},
    xmajorgrids,
    ytick={-0.45, -1.45, -2.45, -3.45, -4.45, -5.45, -6.45, -7.45},
    yticklabel style={align=center},
    yticklabels={{SCM IMEX \\ $\alpha=10^{-12}$}, {SCM CDM \\ $\alpha=10^{-8}$}, {SCM CDM \\ $\alpha=10^{-4}$}, {SCM CDM \\ $\epsilon=10^{-6}$}, {SCM CDM \\ $\epsilon=10^{-4}$}, {IGA-FCM CDM \\ $\alpha=10^{-8}$}, {IGA-FCM CDM \\ $\alpha=10^{-4}$}, {IGA-FCM NM}},
    xmin=0.0,
    xmax=85.0,
    ymax=-0.6,
    ymin=-7.4,
    xbar,
    bar width=5pt,
    xmin=0,
    enlarge y limits,
    height=17cm, width=6.5cm,
    ]
    \addplot[black, fill=black, nodes near coords, every node near coord/.append style={font=\scriptsize}, nodes near coords align={horizontal}] file[] {tikzpictures/data_newest/avg_tim_acc5_p2.dat};
    \addplot[only marks, mark=*, mark size=2pt, black] file[] {tikzpictures/data_newest/all_tim_acc5_p2.dat};
    \addplot[blue, fill=blue, nodes near coords, every node near coord/.append style={font=\scriptsize}, nodes near coords align={horizontal}] file[] {tikzpictures/data_newest/avg_tim_acc5_p3.dat};
    \addplot[only marks, mark=*, mark size=2pt, blue] file[] {tikzpictures/data_newest/all_tim_acc5_p3.dat};
    \addplot[red, fill=red, nodes near coords, every node near coord/.append style={font=\scriptsize}, nodes near coords align={horizontal}] file[] {tikzpictures/data_newest/avg_tim_acc5_p4.dat};
    \addplot[only marks, mark=*, mark size=2pt, red] file[] {tikzpictures/data_newest/all_tim_acc5_p4.dat};
    \addplot[green, fill=green, nodes near coords, every node near coord/.append style={font=\scriptsize}, nodes near coords align={horizontal}] file[] {tikzpictures/data_newest/avg_tim_acc5_p5.dat};
    \addplot[only marks, mark=*, mark size=2pt, green] file[] {tikzpictures/data_newest/all_tim_acc5_p5.dat};

    \end{axis}
\end{tikzpicture}
	\end{subfigure}
    \hspace{2cm}
    \captionsetup[subfigure]{oneside,margin={-2.3cm,0cm}}
	\begin{subfigure}[t]{0.4\textwidth}
		\centering
		\begin{tikzpicture}
	\begin{axis}[
    xlabel={Computation Time $[s]$},
    xmajorgrids,
    ytick={-0.45, -1.45, -2.45, -3.45, -4.45, -5.45, -6.45, -7.45},
    yticklabels={{}, {}, {}, {}, {}, {}, {}, {}, {}},
    xmin=0.0,
    xmax=600.0,
    ymax=-0.6,
    ymin=-7.4,
    xbar,
    bar width=5pt,
    xmin=0,
    enlarge y limits,
    height=17cm, width=6.5cm,
    my xbar legend,
    ]

    \addplot[black, fill=black, nodes near coords, every node near coord/.append style={font=\scriptsize}, nodes near coords align={horizontal}] file[] {tikzpictures/data_newest/avg_tim_acc1_p2.dat};
    \addplot[only marks, mark=*, mark size=2pt, black] file[] {tikzpictures/data_newest/all_tim_acc1_p2.dat};
    
    \addplot[blue, fill=blue, nodes near coords, every node near coord/.append style={font=\scriptsize}, nodes near coords align={horizontal}] file[] {tikzpictures/data_newest/avg_tim_acc1_p3.dat};
    \addplot[only marks, mark=*, mark size=2pt, blue] file[] {tikzpictures/data_newest/all_tim_acc1_p3.dat};
    
    \addplot[red, fill=red, nodes near coords, every node near coord/.append style={font=\scriptsize}, nodes near coords align={horizontal}] file[] {tikzpictures/data_newest/avg_tim_acc1_p4.dat};
    \addplot[only marks, mark=*, mark size=2pt, red] file[] {tikzpictures/data_newest/all_tim_acc1_p4.dat};
    
    \addplot[green, fill=green, nodes near coords, every node near coord/.append style={font=\scriptsize}, nodes near coords align={horizontal}] file[] {tikzpictures/data_newest/avg_tim_acc1_p5.dat};
    \addplot[only marks, mark=*, mark size=2pt, green] file[] {tikzpictures/data_newest/all_tim_acc1_p5.dat};
    \legend{$p = 2$, {}, $p = 3$, {}, $p = 4$, {}, $p = 5$, {}}

    \end{axis}
\end{tikzpicture}
	\end{subfigure}
	\caption{Computation times on Hardware 1: 5\% accuracy (left) 1\% accuracy (right).}
	\label{fig:runtimes_tim}
\end{figure}
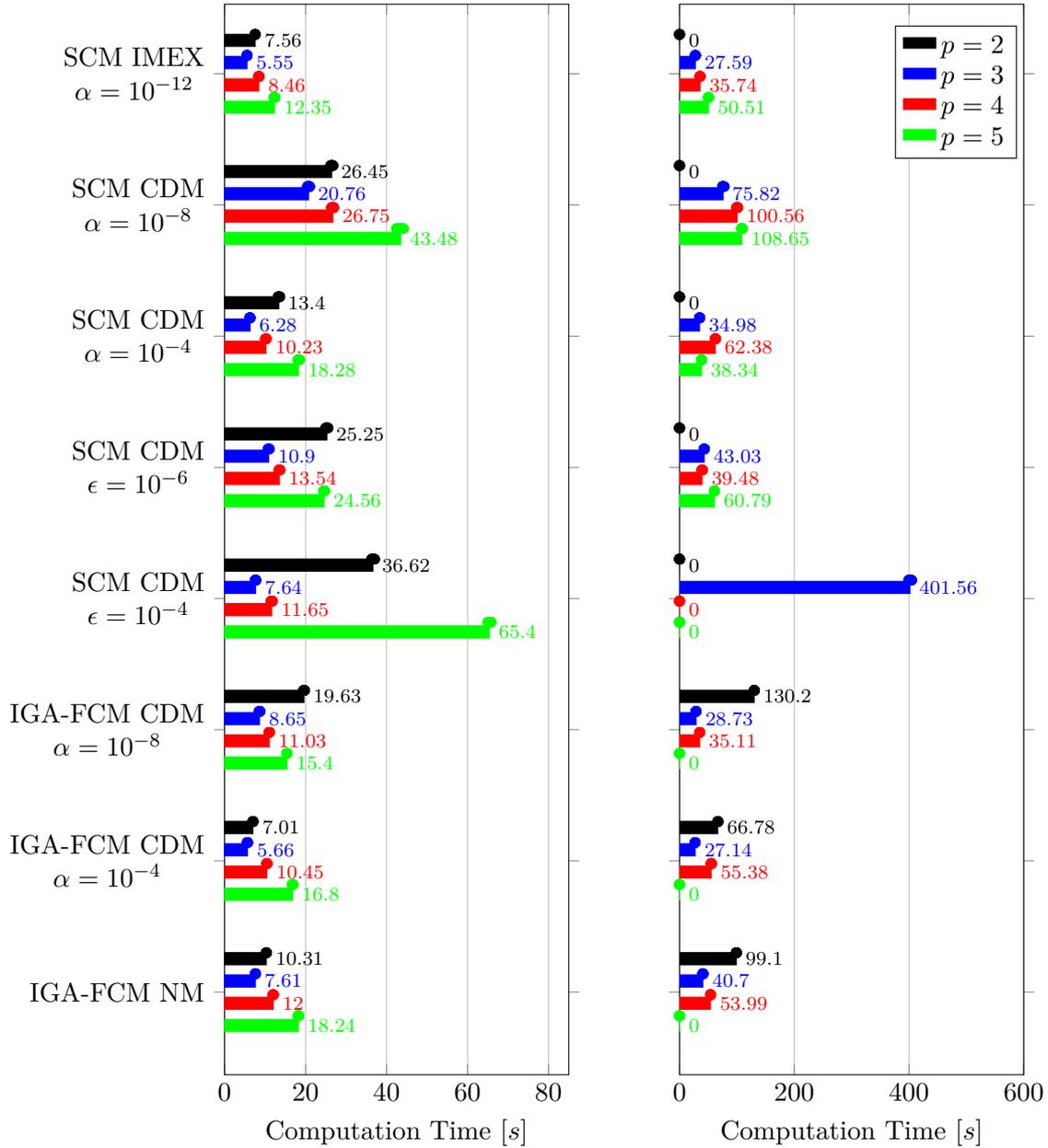
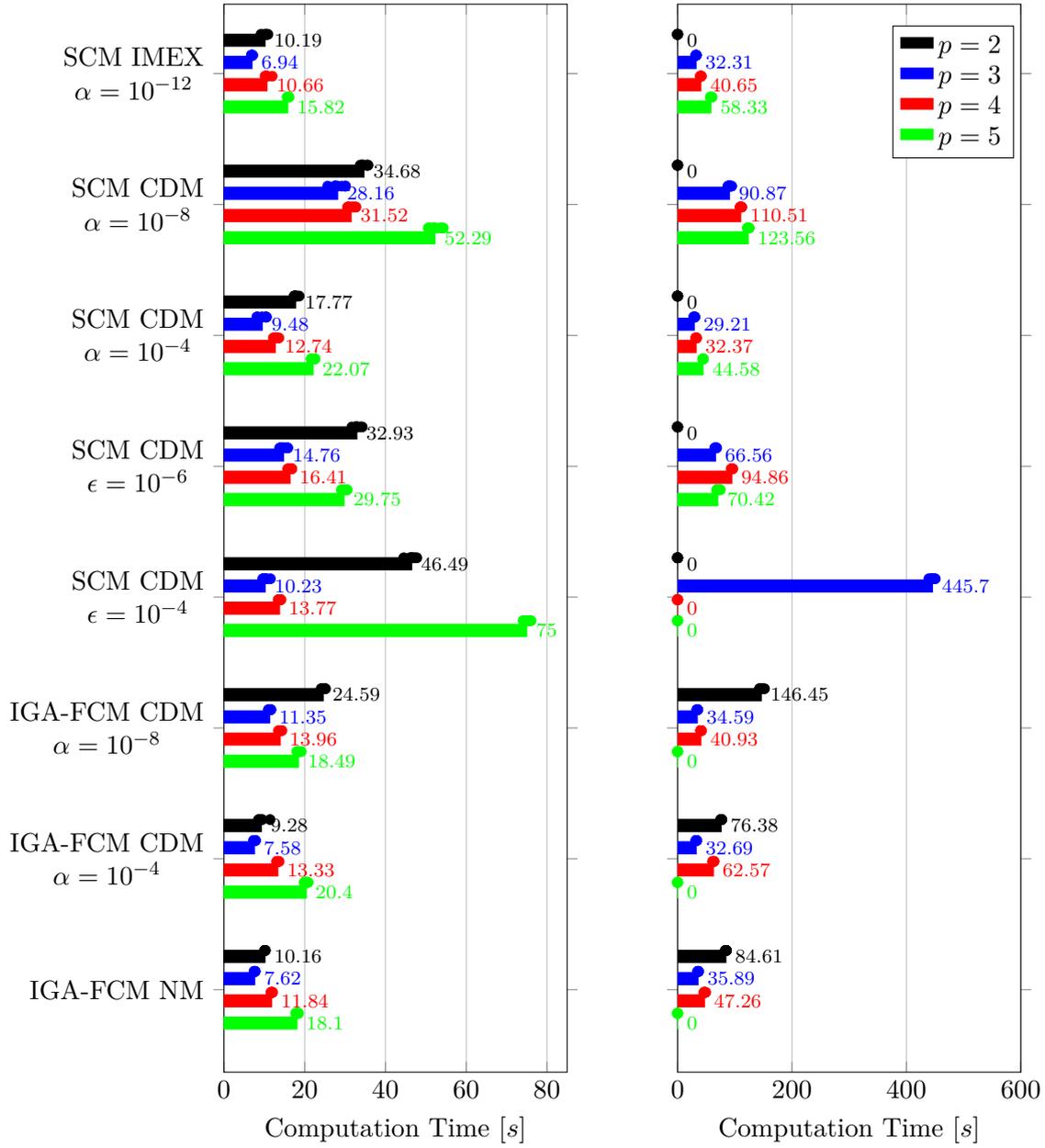
\begin{figure}[htb!]
    \captionsetup[subfigure]{oneside,margin={2.5cm,0cm}}
	\begin{subfigure}[t]{0.4\textwidth}
		\centering
		\begin{tikzpicture}
	\begin{axis}[
    xlabel={Computation Time $[s]$},
    xmajorgrids,
    ytick={-0.45, -1.45, -2.45, -3.45, -4.45, -5.45, -6.45, -7.45},
    yticklabel style={align=center},
    yticklabels={{SCM IMEX \\ $\alpha=10^{-12}$}, {SCM CDM \\ $\alpha=10^{-8}$}, {SCM CDM \\ $\alpha=10^{-4}$}, {SCM CDM \\ $\epsilon=10^{-6}$}, {SCM CDM \\ $\epsilon=10^{-4}$}, {IGA-FCM CDM \\ $\alpha=10^{-8}$}, {IGA-FCM CDM \\ $\alpha=10^{-4}$}, {IGA-FCM NM}},
    xmin=0.0,
    xmax=85.0,
    ymax=-0.6,
    ymin=-7.4,
    xbar,
    bar width=5pt,
    xmin=0,
    enlarge y limits,
    height=17cm, width=6.5cm,
    ]
    \addplot[black, fill=black, nodes near coords, every node near coord/.append style={font=\scriptsize}, nodes near coords align={horizontal}] file[] {tikzpictures/data_newest/avg_lars_acc5_p2.dat};
    \addplot[only marks, mark=*, mark size=2pt, black] file[] {tikzpictures/data_newest/all_lars_acc5_p2.dat};
    \addplot[blue, fill=blue, nodes near coords, every node near coord/.append style={font=\scriptsize}, nodes near coords align={horizontal}] file[] {tikzpictures/data_newest/avg_lars_acc5_p3.dat};
    \addplot[only marks, mark=*, mark size=2pt, blue] file[] {tikzpictures/data_newest/all_lars_acc5_p3.dat};
    \addplot[red, fill=red, nodes near coords, every node near coord/.append style={font=\scriptsize}, nodes near coords align={horizontal}] file[] {tikzpictures/data_newest/avg_lars_acc5_p4.dat};
    \addplot[only marks, mark=*, mark size=2pt, red] file[] {tikzpictures/data_newest/all_lars_acc5_p4.dat};
    \addplot[green, fill=green, nodes near coords, every node near coord/.append style={font=\scriptsize}, nodes near coords align={horizontal}] file[] {tikzpictures/data_newest/avg_lars_acc5_p5.dat};
    \addplot[only marks, mark=*, mark size=2pt, green] file[] {tikzpictures/data_newest/all_lars_acc5_p5.dat};

    \end{axis}
\end{tikzpicture}
	\end{subfigure}
    \hspace{2cm}
    \captionsetup[subfigure]{oneside,margin={-2.3cm,0cm}}
	\begin{subfigure}[t]{0.4\textwidth}
		\centering
		\begin{tikzpicture}
	\begin{axis}[
    xlabel={Computation Time $[s]$},
    xmajorgrids,
    ytick={-0.45, -1.45, -2.45, -3.45, -4.45, -5.45, -6.45, -7.45},
    yticklabels={{}, {}, {}, {}, {}, {}, {}, {}, {}},
    xmin=0.0,
    xmax=600.0,
    ymax=-0.6,
    ymin=-7.4,
    xbar,
    bar width=5pt,
    xmin=0,
    enlarge y limits,
    height=17cm, width=6.5cm,
    my xbar legend,
    ]

    \addplot[black, fill=black, nodes near coords, every node near coord/.append style={font=\scriptsize}, nodes near coords align={horizontal}] file[] {tikzpictures/data_newest/avg_lars_acc1_p2.dat};
    \addplot[only marks, mark=*, mark size=2pt, black] file[] {tikzpictures/data_newest/all_lars_acc1_p2.dat};
    
    \addplot[blue, fill=blue, nodes near coords, every node near coord/.append style={font=\scriptsize}, nodes near coords align={horizontal}] file[] {tikzpictures/data_newest/avg_lars_acc1_p3.dat};
    \addplot[only marks, mark=*, mark size=2pt, blue] file[] {tikzpictures/data_newest/all_lars_acc1_p3.dat};
    
    \addplot[red, fill=red, nodes near coords, every node near coord/.append style={font=\scriptsize}, nodes near coords align={horizontal}] file[] {tikzpictures/data_newest/avg_lars_acc1_p4.dat};
    \addplot[only marks, mark=*, mark size=2pt, red] file[] {tikzpictures/data_newest/all_lars_acc1_p4.dat};
    
    \addplot[green, fill=green, nodes near coords, every node near coord/.append style={font=\scriptsize}, nodes near coords align={horizontal}] file[] {tikzpictures/data_newest/avg_lars_acc1_p5.dat};
    \addplot[only marks, mark=*, mark size=2pt, green] file[] {tikzpictures/data_newest/all_lars_acc1_p5.dat};
    \legend{$p = 2$, {}, $p = 3$, {}, $p = 4$, {}, $p = 5$, {}}

    \end{axis}
\end{tikzpicture}
	\end{subfigure}
	\caption{Computation times on Hardware 2: 5\% accuracy (left) 1\% accuracy (right).}
	\label{fig:runtimes_lars}
\end{figure}

\subsubsection{Discussion}

We can observe that a polynomial degree of $p=3$ is optimal for the majority of methods, taking both accuracy levels into account. 
The sole exception is the SCM with $\epsilon$-stabilization and $\epsilon = 10^{-6}$ for the 1\% error limit on Hardware 1 where $p=4$ delivers the fastest results. 
Nevertheless, at the higher accuracy level, the computation times of polynomial order $p=4$ are getting closer to those of polynomial order $p=3$.
Accordingly, for larger problems with a less detrimental ratio between cut and uncut elements, polynomial degree $p=4$ is also a viable option.
The SCM combined with IMEX time integration and IGA-FCM with a high stabilization by $\alpha = 10^{-4}$ are the fastest methods on both hardwares and for both accuracy levels.
It should be noted that SCM with $\alpha$-stabilization in combination with CDM is also a viable option.
A relatively high value of $\alpha = 10^{-4}$ still permits the desired accuracy levels, and yields computation times that are competitive with those of IMEX time integration.
For Hardware 2, SCM CDM with $\alpha = 10^{-4}$ and $p=3$ even yields the fastest computation times for 1\% accuracy.
For the low accuracy level of 5\%, a high $\epsilon$-stabilization with $\epsilon = 10^{-4}$ also results in satisfactory runtimes. 
However, for higher accuracy, the error introduced by the stabilization becomes too high. 
The use of IGA-FCM is also viable in combination with the trapezoidal Newmark method, which exploits unconditional stability, leading to good runtimes.

\clearpage
\newpage

\section{Conclusion}
\label{sec:conclusion}

In this paper we present the results of efficiency studies conducted for the finite cell method (FCM) when combined with explicit, implicit, and implicit-explicit time integration schemes.
In particular, we determine the required computation time for a given accuracy demand for a variety of different discretization strategies. 
The strategies employ two types of shape functions (Lagrange polynomials and B-splines), resulting in the SCM (spectral cell method) and the IGA-FCM (isogeometric analysis variant of the finite cell method).
The issues related to the critical time step size of such immersed boundary discretizations are addressed using three remedies proposed in previous works:
\begin{enumerate}
    \item The classical material stabilization or $\alpha$-stabilization
    \item The eigenvalue stabilization (EVS)  or $\epsilon$-stabilization
    \item An IMEX (implicit-explicit) time integration scheme
\end{enumerate}
Mass lumping (including cut cells) was considered as well, however, none of the discretization strategies lead to a competitive method when using row-summing or diagonal scaling.
The investigations based on the benchmark problem indicate that three strategies can be recommended in terms of computation time for a given accuracy:
\begin{enumerate}
    \item SCM with mild $\alpha$-stabilization ($\alpha=10^{-12}$) and no $\epsilon$-stabilization ($\epsilon=0$) combined with IMEX.
    \item IGA-FCM with high $\alpha$-stabilization ($\alpha=10^{-4}$) combined with CDM.
    \item SCM with high $\alpha$-stabilization ($\alpha=10^{-4}$) and no $\epsilon$-stabilization ($\epsilon=0$) combined with CDM.
\end{enumerate}
IGA-FCM with with mild $\alpha$-stabilization ($\alpha=10^{-12}$) combined with implicit time integration yields also reasonable computational times. 
SCM with mild $\alpha$-stabilization ($\alpha=10^{-12}$) and a considerable amount of $\epsilon$-stabilization ($\epsilon=10^{-4}$) combined with CDM is also acceptable for low accuracy requirements.
For all of these favorable candidates, a polynomial degree of $p=3$ performs best. 
However, $p=2$ and $p=4$ are almost as efficient, while $p=5$ performs noticeably worse for all candidates. 
Generally, we would like to emphasize that the performance of the methods depends also on the problem under consideration, in particular on the underlying geometry.
The number of cut cells has an significant influence on the efficiency of purely explicit strategies based on the CDM and semi-explicit strategies based on the IMEX scheme. 
In the context of foams, e.g., where the majority of cells are cut, IMEX may be significantly more expensive than indicated here.

Finally, we conclude with some remarks on computational complexity and parallelization.
All results presented are based on solving the underlying linear systems of equations on a single thread with a direct solver (\texttt{pardiso\_64} from Intel's oneMKL), which essentially counts the number of operations to solve the system.
Obviously, for large problems, the use of a direct solver becomes impractical.
Iterative solvers open the possibility of high-performance distributed memory parallel computing.
In such implementations, the method is typically implemented in a matrix-free fashion. 
The number of integration points and the conditioning of the systems become crucial in addition to the size of the system, the support of the basis functions, and the critical time step size.
The question of runtime efficiency of the presented immersed discretization methods in the context of shared and/or distributed memory parallelized implementations remains open for further analysis and discussion.
Nevertheless, the presented results provide representative indications for small- to medium-sized problems.


\appendix


\section*{Acknowledgements} \label{sec:acknowledgement}
{
We gratefully thank the Deutsche Forschungsgemeinschaft (DFG, German Research Foundation) for their support through the grants KO 4570/1-2 and RA 624/29-2 (both grant number 438252876) as well as DU 405/20-1 (grant number 503865803).
}


\bibliographystyle{ieeetr}

\setlength{\bibsep}{3pt}
\setlength{\bibhang}{0.75cm}{\fontsize{9}{9}\selectfont

\end{document}